\documentclass[paper, 12pt, letterpaper]{JHEP} 

\usepackage{amscd,amsmath,amssymb,euscript}
\usepackage[frame,cmtip,curve,arrow,matrix,line,graph]{xy}
\usepackage{amsmath}
\usepackage{amssymb}
\usepackage{graphics}

\def\Bbb{\mathbb} \def\C{{\Bbb C}}  \def\Z{{\Bbb Z}} \def\M{\Bbb{M}}
\def\N {\Bbb{N}}

  \def\P{{\Bbb P}}

\def\gr{{\rm gr}}

\def\Gr{{\rm Gr}}

\def\DG{{\rm DG}}
\def\ord{{\rm ord}}

\def\Hom{\operatorname{Hom}} 
 
    \def\tr{\operatorname{tr}}

\def\Str{\operatorname{Str}}

\def\mod{{\rm mod}}
\def\Gr{{\rm Gr}}
\def\gr{{\rm gr}}

\def\Ext{{\rm Ext}}

\def\cA{{\cal A}}
\def\cT{{\cal T}}

\def\cG{{\cal G}}

\def\cD{{\cal D}}
\def\Ob{{\rm Ob}}

\def\ker{{\rm ker}}  
\def\im{{\rm im\,}}  
 \def\tr{{\rm tr\, }} \def\dim{{\rm dim}}
  
   \def\deg{{\rm deg}}  
  
\def\End{{\rm End}} \def\Aut{{\rm Aut}} \def\Isom{{\rm Isom}}

\def\id{{\rm id}}

\newcommand{\be}{\begin{equation}} \newcommand{\ee}{\end{equation}}
\newcommand{\bea}{\begin{eqnarray}} \newcommand{\eea}{\end{eqnarray}}
\newcommand{\beann}{\begin{eqnarray*}}
  \newcommand{\eeann}{\end{eqnarray*}}
\newcommand{\bfig}{\begin{figure}} \newcommand{\efig}{\end{figure}}
\newcommand{\nn}{\nonumber}
\newcommand{\ba}{\begin{array}}\newcommand{\ea}{\end{array}}

\newtheorem{Proposition}{Proposition}[section]

\newtheorem{Theorem}{Theorem}[section]
\newtheorem{Lemma}{Lemma}[section]
\newtheorem{Corrolary}{Corrolary}[section]

\newcommand{\bp}{\begin{Proposition}}
  \newcommand{\ep}{\end{Proposition}}
\newcommand{\bt}{\begin{Theorem}} \newcommand{\et}{\end{Theorem}}
\newcommand{\bl}{\begin{Lemma}} \newcommand{\el}{\end{Lemma}}
\newcommand{\bc}{\begin{Corrolary}} \newcommand{\ec}{\end{Corrolary}}

   \def\ep{\eps}

\author{~~~C. I.  Lazaroiu\\
\noindent Trinity College Dublin\\
\noindent Dublin 2\\
\noindent Ireland\\
\noindent calin@maths.tcd.ie\\
}

\title{Graded D-branes and skew categories}

\abstract{I describe extended gradings of open topological field
theories in two dimensions in terms of skew categories, proving a
result which alows one to translate between the formalism of graded
open 2d TFTs and equivariant cyclic categories.  As an application of
this formalism, I describe the open 2d TFT of graded D-branes in
Landau-Ginzburg models in terms of an equivariant cyclic structure on the
triangulated category of `graded matrix factorizations' introduced by
Orlov. This leads to a specific conjecture for the Serre functor on
the latter, which generalizes results known from the minimal and
Calabi-Yau cases. I also give a description of the open 2d TFT of such
models which manifestly displays full grading induced by the vector-axial R-symmetry group.}

\preprint{}

\begin{document}

\tableofcontents

\pagebreak

\vskip .6in

\section*{Introduction} 

In open topological string models, the space of
boundary or boundary condition changing observables is generically
$\Z_2$-graded.  In certain cases this grading extends to a larger Abelian group $G$, such as the group of characters of
the surviving vector and axial R-symmetry of the worldsheet model. In
such theories, the category $\cG$ of boundary sectors is $G$-graded,
and the general formalism of two-dimensional open topological field theories
\cite{CIL1,Moore} must be enriched to account for this supplementary structure.  

A basic technical device in the study of open 2d TFTs is the idea of
reduction to an ungraded category, which is implicit in most studies
of homological mirror symmetry. This reduces the structure of
\cite{CIL1, Moore} to a subset of the usual theory of Serre functors
\cite{BK_Serre}.  Such a reduction is possible provided that the
graded category of boundary sectors admits sufficient symmetries which
allow one to trade in the grading for an action on the collection of
objects. In the generic, $\Z_2$-graded case, this translates the
structure of \cite{CIL1} into a Calabi-Yau category whose shift
functor squares to the identity. A similar relation exists when the
open 2d TFT is $\Z_\omega$-graded, the corresponding object being a
Calabi-Yau category whose shift functor has order $\omega$. Finally, a
$\Z$-graded open 2d TFT corresponds to a Calabi-Yau category whose
shift functor generates a $\Z$-subgroup of the automorphism group.

When working with extended gradings by larger Abelian groups $G$, this point
of view leads to the theory of categories endowed with a group action,
and their skew categories \cite{CB}.  In this paper, I discuss the
general framework allowing for such a reduction, giving a precise
relation between the physics description in terms of a $G$-graded
category endowed with natural and nondegenerate traces and a more
familiar mathematical description in terms of `equivariant cyclic
categories', i.e.  ungraded categories endowed with an (Abelian) group action and
a special type of Serre functor.  This leads to a reconstruction
result which allows one to recover the physics description from more
traditional mathematical data. When the group $G$ is sufficiently large, 
this result shows how the formalism of \cite{CIL1, Moore} can 
produce non-Calabi-Yau categories. As an application, I give a complete
treatment of grading issues for the case of topological
Landau-Ginzburg models, refining previous discussion of such theories.

A common example of extended grading in open 2d TFTs arises from open
sigma models which admit an unbroken subgroup $\Gamma$ of vector and
axial R-symmetries. In this situation, the category $\cG$ of boundary
sectors admits a grading by the Abelian group of characters
$G=(\Gamma/\Gamma_0)^*$, where $\Gamma_0$ is the trivially acting
subgroup. The main application considered in this paper concerns
B-type topological Landau-Ginzburg models with target $\C^n$ and bulk
superpotential $W$, whose D-branes were constructed in
\cite{KY,Lerche_LG,LG1,LG2, Walcher}. Using the normalization in which vector
R-charges are integral, the bulk superpotential preserves the full
worldsheet $U(1)_V$ R-symmetry provided that it is a homogeneous
polynomial of the bulk superfields, whose degree we denote by $h=\deg
W$. The model also admits an axial R-symmetry, which is broken to a
$\Z_2$ subgroup by the bulk superpotential. This is responsible for
the appearance of supermodules in the description of
\cite{KY,Lerche_LG,LG1,LG2} and gives an unbroken total R-symmetry
group $\Gamma=U(1)\times \Z_2$.  It turns out that the grading group $G$ depends on the parity of $h$. In
fact, direct analysis of the worldsheet shows that the trivially
acting subgroup $\Gamma_0$ is trivial for even $h$ and isomorphic to
$\Z_2$ for odd $h$. As a consequence, $G$ equals
$\Z\times \Z_2$ or $\Z$, depending on the parity of $h$. Taking this
fact into account, one finds that the category of `graded D-branes'
(i.e. those B-type branes which preserve the full vector R-symmetry)
is most conveniently described as the category of finitely generated
$G$-graded integrable modules over a curved differential graded
algebra with trivial differential. This simplifies the more
traditional description in terms of matrix factorizations, and allows
one to give a complete construction of the differential graded
category underlying graded D-branes, in a formulation which 
explicitly displays the worldsheet symmetry of the model. The formulation 
used in this paper is a Landau-Ginzburg extension of the approach through graded superconnections 
of total degree one used in \cite{CIL2, CIL3, CIL4, Diac, CIL6, CIL7, CIL8, Block} for B-type branes on 
Calabi-Yau manifolds.
 
Using this description, one finds that the resulting open 2d TFT is described by a `$G$-graded
category with shifts', whose shift functors are compatible with the
invariant traces. Restricting to morphisms of trivial degree gives an
equivalent description in terms of a $G$-equivariant cyclic structure
on the triangulated category of graded matrix factorizations
constructed in \cite{Orlov3}. In particular, we find that the traces of
\cite{KY} lead to a precise conjecture for a Serre functor on this
triangulated category. Finally, the correspondence result mentioned
above allows one to recover the full open 2d TFT from this triangulated category, 
provided that the latter is considered together with this Serre functor.

The paper is organized as follows. In Section 1, we discuss the
mathematical notion of open 2d TFTs with extended grading, and explain
how examples of such gradings arise from worldsheet
R-symmetries. In Section 2, we discuss the special case of `graded
open 2d TFTs with shifts', which allows for an equivalent description
in terms of ungraded categories endowed with an equivariant cyclic
structure. After developing some mathematical machinery, we give a
result which translates between the two descriptions. This gives a
precise relation between the nondegenerate traces required by the
physics formalism and the usual theory of Serre functors
\cite{BK_Serre}. Section 3 discusses dG categories with extended
grading, focusing on the example of integrable modules with extended
grading over a curved dGA, which is relevant for Landau-Ginzburg
models. In Section 4, we apply this machinery to the case of
Landau-Ginzburg models with target $\C^n$. Using the worldsheet
analysis Appendix \ref{sec:worldsheet}, we show that graded matrix factorizations
can be described as the $G$-graded dG category of integrable modules
over a curved differential graded algebra, whose total cohomology
category recovers the desired $G$-graded open 2d TFT with
shifts. After giving the precise homogeneity property of the traces, 
we use the general correspondence of Section 2 to
extract a conjecture for the Serre functor on the triangulated
category of matrix factorizations, and check it against known results in the Calabi-Yau 
and minimal model case \cite{Takahashi,saito}. This fills in a gap in the relation
between the mathematical analysis of \cite{Orlov3} and the physical
description of open 2d TFTs. Appendix \ref{sec:Serre} 
contains a general discussion of equivariant Serre functors, while Appendix \ref{sec:worldsheet} 
gives the worldsheet analysis which leads to the description used in Section 4.

\section{Graded open topological field theories in two dimensions}
\label{sec:graded_TFT}

It was shown in \cite{CIL1, Moore} that the boundary sector of a topological
field theory in two dimensions is encoded by category-theoretic data,
which can be viewed as an abstract definition of the notion of open 2d
topological field theory (TFT).  As shown in loc. cit., such a
theory is described by a Hom-finite\footnote{Recall that $\cG$ is called Hom-finite if $\dim_\C\Hom_\cG(a,b)<\infty$ for all 
$a,b\in \Ob\cG$.} $\C$-category $\cG$, whose objects are the
boundary sectors, and whose morphisms are the boundary/boundary
condition changing observables; the morphism compositions correspond to
the product of such observables.  A distinguishing feature of such
theories is the existence of nondegenerate invariant linear maps
$\tr_a:\Hom_{\cG}(a,a)\rightarrow \C$, which arise from the two-point
functions on the disk. In the generic case, the category
$\cG$ is $\Z_2$-graded, while the traces are graded-symmetric and $\Z_2$-homogeneous of a
common degree which is a characteristic of the model. 

It is often the case that $\cG$ carries a grading by some larger Abelian group  $G$, which we will write in additive notation. 
The traces of such models are homogeneous of some common degree $-\kappa\in G$, and must satisfy a graded-symmetry
condition whose signs involve a group morphism $\epsilon:G\rightarrow
\Z_2$. In fact, $G$ can be viewed as an extension of $\Z_2$ by the kernel of $\epsilon$.
To capture this, the formalism of \cite{CIL1, Moore} can be enhanced as follows. 
Recall that a $\C$-category $\cG$ is called $G$-graded 
if we are given direct sum decompositions:
\be
\label{Hom_decomp}
\Hom_{\cG}(a,b)=\oplus_{g\in G}\Hom_{\cG}^g(a,b)~~
\ee
for all $a,b\in \Ob \cG$ such the following conditions are satisfied:

1) We have $\id_a\in \Hom^0_{\cG}(a,a)$ for all objects $a$ of $\cG$.

2) We have $\Hom_{\cG}^h(b,c)\circ \Hom_{\cG}^g(a,b)\subset
   \Hom_{\cG}^{h+g}(a,c)$ for all objects $a,b,c$ and all $g,h\in G$.

\noindent Given $u\in \Hom^g_{\cG}(a,b)$, we set $\deg u:=g$. Then models with a boundary space grading by $G$ can be described
as follows:

\paragraph{Definition}{
Let $G$ be an Abelian group, $\epsilon:G\rightarrow \Z_2$ a group morphism and
$\kappa\in G$ an element of $G$.  A {\em graded open 2d TFT} of type
$(G,\kappa,\epsilon)$ is a pair $(\cG, \tr)$ where $\cG$ is a
Hom-finite $G$-graded $\C$-category and $\tr$ is a family of maps
$\tr_a:\Hom_\cG(a,a)\rightarrow \C$ defined for all $a\in \Ob\cG$,
such that:

(1) The following selection rule is satisfied for any homogeneous $u\in \Hom_\cG(a,a)$:
\be
\label{tr_sel}
\tr_a(u)=0~~{\rm unless}~~\deg u=\kappa~~.
\ee

(2) The following conditions are satisfied for all $a,b\in \Ob\cG$:
\be 
\label{tr_sym}
\tr_a(vu)=(-1)^{\epsilon(\deg u)\epsilon(\deg
v)}\tr_b(uv)~~\forall ~~{\rm homogeneous}~u\in \Hom_{\cG}(a,b)~~{\rm
and}~~v\in \Hom_{\cG}(b,a)~~.  \ee

(3) For any objects $a,b$ of $\cG$ and any $u\in \Hom_{\cG}(a,b)$, the
following implication holds: 
\be 
\label{tr_nondeg}
\tr_a(vu)=0~~\forall v\in
\Hom_{\cG}(b,a)\Rightarrow u=0~~.  
\ee

Trivial examples are provided by generic open 2d TFTs, with $G=\Z_2$ and $\epsilon=\id_{\Z_2}$. 
More complicated models 
arise by considering an extended worldsheet symmetry group $\Gamma$ which is preserved by the
boundary conditions corresponding to the objects of $\cG$.  Then $\Gamma$
acts on the spaces of boundary and boundary condition-changing
observables, i.e. it acts by automorphisms of $\cG$ which fix all of its objects. If $\Gamma_0$ is the trivially acting subgroup, 
we have representations of $\Gamma/\Gamma_0$ on the
vector spaces $\Hom_{\cG}(a,b)$. Thus we can view $\cG$ as a
$\C[\Gamma/\Gamma_0]$-category, where $\C[\Gamma/\Gamma_0]$ is the
group algebra of $\Gamma/\Gamma_0$. This means that
$\Hom_{\cG}(a,b)$ are $\C[\Gamma/\Gamma_0]$-modules and the
compositions of $\cG$ are $\C[\Gamma/\Gamma_0]$-linear. Since the quotient 
group $\Gamma/\Gamma_0$ is Abelian, the spaces $\Hom_{\cG}(a,b)$ decompose into one-dimensional
irreducible representations, which can be collated together to give decompositions
(\ref{Hom_decomp}), where the grading group $G:=(\Gamma/\Gamma_0)^*$ is the character
group (Pontryaghin dual) of $\Gamma/\Gamma_0$. Compatibility
of the representations with morphism compositions translates into
relations 2) above. On the other hand, the traces satisfy
(\ref{tr_sel}), where $\kappa\in G$ is dictated by the anomaly of the
worldsheet path integral under the action of $\Gamma/\Gamma_0$ (the
later arises because the action of $\Gamma$ typically involves some
nontrivial axial transformation of the worldsheet fermions).

A common application of the above concerns the vector and axial R-symmetries of a
sigma model. Recall that two dimensional sigma models with bulk
$N=(2,2)$ supersymmetry admit two classes of boundary conditions
preserving two supersymmetries, namely boundary conditions of $A$ and $B$-type. 
Each of these types preserves a certain subgroup of the
R-symmetry group of the bulk theory.  At classical level and in the
absence of a bulk superpotential, the bulk theory has vector and axial
R-symmetry groups $U(1)_V$ and $U(1)_A$, each of which can be broken
to subgroups already in the bulk.  For example, $U(1)_A$ is subject to
the quantum anomaly and can also be broken classically (to a $\Z_2$
subgroup) by a bulk superpotential.  On the other hand, $U(1)_V$ is
preserved by homogeneous bulk superpotentials but is broken by
inhomogeneous ones.  Let us denote the surviving vector and axial
R-symmetry subgroups of the {\em quantum} bulk theory by $R_V$ and
$R_A$.  Then A-type boundary conditions preserve\footnote{For the $A$
model, $R_A$ can also suffer a quantum {\em boundary} anomaly, so one must chose
the A-type boundary conditions such that this anomaly vanishes. This
corresponds to the vanishing Maslow index condition in \cite{FOOO}.}
$R_A$ while breaking $R_V$ to a subgroup $R'_V$.  The B-type boundary
conditions preserve $R_V$, while breaking $R_A$ to a subgroup $R_A'$.
Hence the boundary model has a global $R$-symmetry
$\Gamma=\Gamma_A\times \Gamma_V$ where $\Gamma_A=R_A$ and
$\Gamma_V=R'_V$ for A-branes, and $\Gamma_A=R'_A, \Gamma_V=R_V$ for
$B$-branes. As a consequence, the space of boundary observables
carries a representation of the Abelian group $\Gamma$. Define
$\Gamma'=\Gamma/\Gamma_0$, where $\Gamma_0$ is the trivially acting
subgroup.  Then the spaces of boundary/boundary condition changing
observables carry gradings by the Abelian group of characters
$G=(\Gamma')^*$. Upon performing the topological twist, the A (respectively B) -type
branes of the untwisted theory become topological D-branes of the A
(respectively B) model. Then $\Gamma$ acts on the space of
topological observables, which therefore carries a grading by $G$. 
One thus finds that the corresponding open 2d TFT is
$G$-graded.

\section{Graded open 2d TFTs with shifts}
\label{sec:shifts}

In this section, we fix an Abelian group $G$. 

\subsection{Mathematical preparations}

\paragraph{$G$-categories.}

A {\em $G$-category} is a triple
$(\cT,G,\gamma)$ where $G$ is a group, $\cT$ is a $\C$-category and
$\gamma$ is a {\em faithful~}\footnote{We restrict to faithful actions in order to avoid pathologies below.} 
$G$-action on $\cT$, i.e.  a group
monomorphism $\gamma: G\rightarrow \Aut(\cT)$, where $\Aut(\cT)$ is
the group of automorphisms of $\cT$.  Given two $G$-categories, 
a {\em $G$-invariant functor}
$(\cT,G,\gamma)\rightarrow (\cT',G,\gamma')$ is a
functor $F:\cT\rightarrow \cT'$ such that $F\circ
\gamma(g)=\gamma'(g)\circ F$ for all $g\in G$. Small
$G$-categories form a category ${\rm Cat}_G$ whose
morphisms are given by $G$-invariant functors.

\paragraph{Graded categories with shifts}

Given a $G$-graded category $\cG$, we let $\Aut^0(\cG)$ denote the group of degree zero automorphisms, i.e. those automorphisms $F$ of 
$\cG$ such that $F(\Hom_\cG^g(a,b))\subset \Hom_{\cG}^g(a,b)$ for all
objects $a,b$ and all $g\in G$. 

A {\em graded category with shifts} is a quadruple 
$(\cG,G,\gamma,s)$ where $\cG$ is a $G$-graded category, $\gamma:G\rightarrow \Aut^0(\cG)$ 
is a group monomorphism and $s(g):\id_{\cG}\stackrel{\sim}{\rightarrow}\gamma(g)$ are
isomorphisms of functors, subject to the conditions:
\be
\label{sdeg}
~s_a(g)\in \Hom_{\cG}^{-g}(a,\gamma(g)(a))~\ee
and:
\be
\label{sdef}
\gamma(g)(s_{a}(h))\circ s_a(g)=s_a(g+h)~~.
\ee
for all $a, b\in \Ob \cG$ and $g,h \in G$. 

\

\noindent We say that $\gamma(g)$, $s(g)$ are the {\em shifts} and {\em suspensions} of $\cG$. 
Naturality of $s(g)$ reads: 
\be
\label{smf}
\gamma(g)(u)\circ s_a(g)=s_b(g)\circ u~~
\ee
for all $u\in\Hom_{\cG}(a,b)$. Combined with (\ref{sdef}), this gives: 
\be
s_a(g+h)=s_{\gamma(h)(a)}(g)\circ s_a(h)~~,
\ee 
which in turn implies $s_a(e)=\id_a$ and $s_a(-g)=s_{\gamma(-g)(a)}(g)^{-1}$. Using these 
equations, one checks the relation:
\be
\label{sconj}
s_{\gamma(g)(a)}(h)=\gamma(g)(s_a(h))~~,
\ee
which will be useful below. 

A morphism $(\cG,G,\gamma,s)\rightarrow (\cG',G,\gamma', s')$ of
$G$-graded categories with shifts is a functor $F:\cG\rightarrow \cG'$
which preserves morphism degrees, intertwines $\gamma(g)$ and $\gamma'(g)$
and maps $s_a(g)$ into $s'_{F(a)}(g)$ and
intertwines $\gamma(g)$ and $\gamma'(g)$ (the second condition means that
$F:(\cG,G,\gamma)\rightarrow (\cG',G,\gamma')$ is a $G$-invariant functor).  
Small $G$-graded categories with shifts form a 
category ${\rm GrCats}_G$ when endowed with the morphisms given by such
functors.

\paragraph{The skew category of a $G$-category.}

The {\em graded skew category}  of a $G$-category $(\cT,G,\gamma)$ is the
$G$-graded category $\cT^\bullet [G]$ having the same objects as $\cT$, morphism
spaces: 
\be
\label{hd}
\Hom_{\cT^\bullet [G]}(a,b)\stackrel{\rm def}{=}\oplus_{g\in G}\Hom^g_{\cT^\bullet[G]}(a,b)~~
\ee
with :
\be
\label{orbit_def}
\Hom^g_{\cT^\bullet [G]}(a,b)\stackrel{\rm def}{=}\Hom_{\cT}(a,\gamma(g)(b))
\ee
and morphism compositions given by: 
\be
\Hom_{{\cT}^\bullet[G]}^h(b, c)\times \Hom_{{\cT}^\bullet[G]}^g (a, b)\ni (v,u)\longrightarrow v*u:= \gamma(g)(v)\circ u \in \Hom_{{\cT}^\bullet[G]}^{h+g}(a,b)
\ee
The {\em skew category} \cite{CB} $\cT[G]$ of $(\cT,G,\gamma)$ is the
category obtained from $\cT^\bullet[G]$ by forgetting the
$G$-grading.  When the action of $G$ on $\Ob\cT$ is free, $\cT[G]$
is equivalent \cite{CB} with the naive quotient category of $\cT$ by $G$.

The fundamental property of $\cT[G]$ is that objects belonging to the
same $G$-orbit of $\cT$ become isomorphic in $\cT[G]$. Indeed,
$\id_a\in \Hom_{\cT}(a,a)=\Hom_{\cT^\bullet
[G]}^{-g}(a,\gamma(g)(a))$ and $\id_{\gamma(g)(a)}\in
\Hom_{\cT}(\gamma(g)(a),\gamma(g)(a))\approx \Hom_{\cT^\bullet
[G]}^g(\gamma(g)(a),a)$ provide mutually inverse isomorphisms between
$a$ and $\gamma(g)(a)$. We let $s_a(g):a\rightarrow \gamma(g)(a)$
denote the morphism $\id_a$ when viewed as an element of
$\Hom_{\cT^\bullet[G]}^{-g}(a,\gamma(g)(a))$.

Notice that $\gamma(g)$ can be viewed as degree zero automorphisms
$\gamma^\bullet(g)$ of $\cT^\bullet [G]$, giving a $G$-category structure on the latter. 
The maps $s_a(g)$ give isomorphisms of functors
$s(g):\id_{\cT[G]}\stackrel{\sim}{\rightarrow}\gamma(g)$, which
satisfy $\gamma(g)(s(h))\circ s(g)=s(gh)$ for all $h,g\in G$.  Thus
$(\cT^\bullet[G],G, \gamma^\bullet,s)$ is a graded category with
shifts.

\paragraph{Equivariant functors and framings.}

Given two $G$-categories, $(\cT,G,\gamma)$ and $(\cT',G,\gamma')$, 
a functor $F:\cT\rightarrow \cT'$ is called {\em $G$-equivariant} if there
exist isomorphisms of functors $\eta(g):F\circ
\gamma(g)\stackrel{\sim}{\rightarrow} \gamma'(g)\circ F$ satisfying the
compatibility conditions:
\be
\label{eta_comp}
\eta(g_1+g_2)=\gamma(g_1)(\eta(g_2))\circ \eta(g_1)~~,
\ee
where we use the obvious slight abuse of notation. These conditions imply that $\eta(0)$ is the identity 
endomorphism of $F$. Any choice of
$\eta$ satisfying (\ref{eta_comp}) will be called a {\em
framing} of $F$. A framed $G$-equivariant endofunctor $F:\cT\rightarrow \cT'$ induces
an endofunctor $F^\bullet:\cT^\bullet[G]\rightarrow \cT'^\bullet[G]$ given by $F^\bullet (a):=F(a)$ on
objects and acting on morphisms as follows:
\be
\nn
\!\!\!\!\!\!\! u\in \Hom^g_{\cT^\bullet[G]}(a,b)=\Hom_{\cT}(a,\gamma(g)(b)) \rightarrow F^\bullet_{ab}(u):=
\eta_b(g)\circ F_{a \gamma(g)(b)}(u) \in \Hom^g_{\cT'^\bullet [G]}(F(a),F(b))~~.
\ee
Notice that $F^\bullet$ preserves the degree of morphisms. 

Obviously, a $G$-invariant functor is $G$-equivariant and admits the
trivial framing given by $\eta(g)=\id$ for all $g$. Given an invariant
functor $F$, we make the convention that $F^\bullet$ denotes the
functor induced by the trivial framing. Since $G$ is Abelian, each
$\gamma(g)$ is $G$-invariant and trivially framed, thus inducing the endofunctor
$\gamma^\bullet(g)$ of $\cT^\bullet[G]$ mentioned above.

Given framed $G$-equivariant endofunctors $(F,\eta^F)$ and $(G, \eta^G)$, their composition
$F\circ G$ is $G$-equivariant and admits the {\em composite framing}
$\eta^{F\circ G}(g)=\eta^F(g)\circ F(\eta^G(g))$.  We have
$(F\circ G)^\bullet=F^\bullet\circ G^\bullet $ when $F\circ G$ is
endowed with this framing. Applying this to $G$-invariant functors (which we take to be trivially framed), we find
that $^\bullet$ gives a functor form ${\rm Cat}_G$ to ${\rm GrCats}_G$.

\paragraph{Orbit categories}

Given an automorphism $\sigma$ of $\cT$, let $G_\sigma$ be the
subgroup of $\Aut(\cT)$ generated by $\sigma$, which acts tautologically on $\cT$ via
the inclusion $G_\sigma\hookrightarrow \Aut(\cT)$. We define the {\em
orbit category} \cite{Keller_orbit} of $\cT$ modulo $\sigma$ to be the skew category
$\cT_\sigma\stackrel{\rm def}{=}\cT[G_\sigma]$. We have
$G_\sigma\approx \Z_\omega:=\Z/\omega\Z$, where $\omega =\ord \sigma$ (we set $\omega=0$ when no power of $\sigma$ equals the identity functor).  Therefore,
$\cT_\sigma^\bullet:=\cT^\bullet[G_\sigma]$ is $\Z_\omega$-graded via $\Hom_{{\cT_\sigma}^\bullet}^n(a,b)=
\Hom_{\cT}(a,\sigma^n(b))$ for all $n\in \Z_\omega$.
An endofunctor $F$ of $\cT$ is
$G_\sigma$-equivariant iff $F\circ \sigma\approx \sigma\circ F$, with
strict equality when $F$ is $G_\sigma$-invariant. 

Notice that $\sigma$ is $G_\sigma$-invariant, thus inducing an
automorphism $\sigma^\bullet$ of $\cT^\bullet_\sigma$ when endowed with the
trivial framing. We have $\sigma^\bullet \approx \id_{\cT_\sigma}$ in
$\cT_\sigma$, via the isomorphism of functors given by $\id_a\in
\Isom^{-1}_{\cT^\bullet [G]}(a, \sigma(a))$.  Moreover, we have
$\Hom_{{\cT_\sigma}^\bullet}^n(a, \sigma (b))=\Hom^{n+1}_{{\cT_\sigma}^\bullet}(a,b)$,
which shows that $\sigma^\bullet$ is a shift functor for the
$\Z_\omega$-graded category $\cT_\sigma^\bullet$. The application of orbit categories to open 
2d TFTs  was discussed in some detail in Appendix A of \cite{gen}.

\paragraph{The null restriction of a graded category with shifts}

Given a graded category  with shifts $(\cG,G,\gamma,s)$, the
non-full subcategory $\cG^0$ obtained from $\cG$ by keeping all objects but restricting to
morphisms of trivial degree carries the $G$-action $\gamma^0$ obtained by
Co-restricting $\gamma$.  Hence 
$(\cG^0,G,\gamma^0)$ is a $G$-category.  Given a morphism
$F:(\cG,G,\gamma,s)\rightarrow (\cG',G,\gamma',s')$ in ${\rm GrCats}_G$, 
we let $F^0:(\cG^0,G,\gamma^0)\rightarrow (\cG'^0,G,\gamma'^0)$ be the morphism in ${\rm Cat}_G$
obtained by restricting the action $F$ to morphisms of degree zero. 
This gives a functor $^0:{\rm GrCats}_G\rightarrow {\rm Cat}_G$.

\paragraph{Relation between skew categories and graded categories with shifts}
\label{sec:relation}

The following result characterizes skew categories as graded categories with shifts.

\paragraph{Proposition} The functors $^\bullet:{\rm Cat}_G\rightarrow {\rm
GrCats_G}$ and $^0:{\rm GrCats}_G\rightarrow {\rm Cat}_G$ are mutually quasi-inverse
equivalences.

\

\noindent{\em Proof.} (1) Given a $G$-category $(\cT,G,\gamma)$, we
showed above that $(\cT^\bullet[G],G,\gamma^\bullet,s)$ is a $G$-graded category
with shifts. Setting $\cG:=\cT^\bullet[G]$, it is clear that $\cG^0$
coincides with $\cT$ as a $G$-category. Given a morphism $F$ in ${\rm
Cat}_G$, it is also obvious that $(F^\bullet)^0=F$.  Hence the
composition $[(-)^\bullet]^0$ is the identity functor of ${\rm Cat}_G$.

(2) Given a $G$-graded category with shifts $(\cG,G,\gamma,s)$,
consider the $G$-category $(\cT,G,\gamma^0)$ where $\cT:=\cG^0$. We
define a functor of graded categories $\Phi:=\Phi_\cG:\cG\rightarrow
\cT^\bullet[G]$ which acts trivially on objects (i.e.  $\Phi(a):=a$
for all $a\in \Ob \cG$) and acts on morphisms as follows:
\be
\nn
\Hom_{\cG}(a,b)\ni u=\oplus_{g\in g}{u_g}\rightarrow \Phi_{ab}(u):=\oplus_{g\in G}{s_b(g)\circ u_g}\in \Hom_{\cT^\bullet[G]}(a,b)~~.
\ee 
Here $u_g\in \Hom_{\cG}^g(a,b)$, so $s_b(g)\circ u_g\in \Hom_{\cG}^0(a,\gamma(g)(b))=\Hom_{\cT}(a,\gamma(g)(b))=\Hom_{\cT^\bullet[G]}^{g}(a,b)$. 
To check that $\Phi$ is a functor, 
pick $u\in \Hom_{\cG}^g(a,b)$ and $v\in \Hom_{\cG}^h(b,c)$ and compute: 
\bea
\Phi_{bc}(v)*\Phi_{ab}(u)=\gamma(g)(s_c(h)\circ v)\circ s_b(g)\circ u=\gamma(g)(s_c(h))\circ \gamma(g)(v)\circ s_b(g)\circ u\nn\\
=\gamma(g)(s_c(h))\circ s_c(g) \circ v\circ u=s_c(g+h)\circ v\circ u=\Phi_{ac}(v\circ u)~~,\nn
\eea
where we used (\ref{smf}) and (\ref{sdef}). Since $\Phi_{ab}$ are
clearly bijective, we find that $\Phi$ is an isomorphism between $\cG$
and $\cT^\bullet[G]=(\cG^0)^\bullet [G]$. 

We now show that $\Phi$ is an isomorphism of graded categories with shifts. 
To show that $\Phi$ commutes
with the $G$-actions, pick $u\in \Hom_{\cG}^h(a,b)$ and compute:
\be
\nn
\Phi(\gamma^\bullet (g)(u))=s_{\gamma(g)(b)}(h)\circ \gamma^\bullet (g)(u)=\gamma(g)(s_b(h))\circ \gamma^\bullet (g)(u)=\gamma(g)(\Phi(u))~~,
\ee 
where we used (\ref{sconj}). Finally, we have $\Phi(s_a(g))=s_{\gamma(g)(a)}(-g)\circ s_a(g)=s_a(e)=\id_a$, where $\id_a$ is viewed as 
an element of $\Hom_{\cT^\bullet[G]}^{-g}(a,\gamma(g)(a))$ as required by  
the definition of $\Phi$.  Thus $\Phi(s_a(g))=s_a^{\cT^\bullet[G]}(g)$ is the corresponding suspension of $\cT^\bullet[G]$. 
We conclude that $\Phi$ is an isomorphism of graded categories 
with shifts. Given a morphism $F:(\cG,G,\gamma,s)\rightarrow (\cG',G,\gamma',s')$ in
${\rm GrCats}_G$, it is easy to check the identity $\Phi_{\cG'}\circ
F=(F^0)^\bullet \circ \Phi_\cG$, which shows that $\Phi_\cG$ give an isomorphism of functors 
$\id_{{\rm Gr Cats}_G}\rightarrow [(-)^0]^\bullet$.

The conclusion now follows by combining (1) and (2). 

\

\noindent The proposition allows us to reconstruct a graded category with shifts $\cG$ as the skew category of its null restriction $\cG^0$, thereby reducing 
the theory of graded categories with shifts to that of $G$-categories. Notice that the functor $\Phi_\cG$ realizing this correspondence is determined 
by the suspensions of $\cG$. In the next subsection, we show how this result can be used to express 
the data of certain graded open 2d TFTs in terms of $G$-categories with supplementary structure.

\paragraph{Convention} In order to simplify notation, we will often use the same 
symbol for the $G$-representation $\gamma$ of a  $G$-category $\cT$ and the 
representation $\gamma^\bullet$ of its skew category $\cT^\bullet[G]$. Which one 
is meant should be clear from the context.

\subsection{$G$-graded open 2d TFTs with shifts}
\label{sec:gradedTFT}

Let $G$ be an Abelian group as before. $G$-graded open 2d topological
field theories can be described through ungraded categories provided
that they admit a $G$-action with special properties. Such special
theories can be described as follows.

\paragraph{Definition} A {\em graded open 2d TFT with shifts} is a system $(\cG, G, \gamma,s,\tr, \kappa,\epsilon)$ such that: 

(1) $(\cG, G, \gamma,s)$ is a Hom-finite $G$-graded $\C$-category with shifts.

(2) $(\cG,\tr)$ is a graded 2d open TFT of type $(G, \kappa,\epsilon)$, where $\kappa\in G$ and $\epsilon \in \Hom(G,\Z_2)$ 

(3) The traces $\tr_a:\Hom_{\cG}(a,a)\rightarrow \C$ satisfy the equivariance conditions:
\be
\label{Equiv}
\tr_{\gamma(g)(a)}(\gamma(g)(u))=(-1)^{\epsilon(g)(\epsilon(\kappa)+1)}\tr_a(u)
\ee
for all $a\in \Ob\cG$, $u\in \Hom_{\cG}(a,a)$ and $g\in G$.

\

\noindent We will see in a moment that this structure is equivalent with the one given in the definition below. 

\paragraph{Definition} Let $(\cT, G, \gamma)$ be a Hom-finite $G$-category and fix an element $\kappa\in G$ and a group morphism 
$\epsilon:G\rightarrow \Z_2$. A {\em G-equivariant cyclic structure} of type $(\kappa,\epsilon)$ on $(\cT, G, \gamma)$ 
is a family of linear maps $tr_a:\Hom_{\cT}(a, \gamma(\kappa)(a))\rightarrow \C$ defined for all $a\in \Ob \cT$ which 
obey the symmetry conditions: 
\be
\label{tr_cycl}
tr_{a}(v\circ u)=tr_b(\gamma(\kappa)(u)\circ v)~~\forall u\in \Hom_{\cT}(a,b)~~\forall v\in \Hom_{\cT}(b,\gamma(\kappa)(a))~~,
\ee
the equivariance conditions: 
\be
\label{tr_equiv}
tr_{\gamma(g)(a)}(\gamma(g)(u))=(-1)^{\epsilon(g)(\epsilon(\kappa)+1)}tr_a(u)~~\forall u\in \Hom_{\cT}(a,\gamma(\kappa)(a))
\ee
and the following nondegeneracy condition for all $u\in \Hom_{\cT}(a,b)$:
\be
\label{nondegen}
tr_a(v\circ u)=0~~\forall v\in\Hom_{\cT}(b,\gamma(k)(a))~\Rightarrow u=0~~.
\ee 
We also say that $(\cT, G, \gamma,tr,\kappa,\epsilon)$ is an {\em equivariant cyclic category} of type $(G,\gamma,\kappa,\epsilon)$. 

\paragraph{Observation} Let  $(\cT, G, \gamma,tr,\kappa,\epsilon)$ be an equivariant cyclic category and set $S:=\gamma(\kappa)$. 
Using the traces $tr_a$, define linear maps $\phi_{ab}:\Hom_{\cT}(a,b)\rightarrow \Hom_{\cT}(b,S(a))^{\rm v}$ via: 
$\phi_{ab}(u)(v):=tr_a(v\circ u)$. Then the pair $(S,\phi)$ is a Serre functor on $\cT$ in the sense of \cite{BK_Serre}. 
Indeed, the symmetry condition (\ref{tr_cycl}) amounts to naturality of  $\phi_{ab}$ in $a$ and $b$, while the nondegeneracy 
condition (\ref{nondegen}) amounts to bijectivity of $\phi_{ab}$. The traces can be recovered as $\tr_a=\phi_{aa}(\id_a)$. 
Since $G$ is Abelian, the functor $S=\gamma(\kappa)$ is $G$-invariant. On the other hand, condition (\ref{tr_equiv}) becomes:
\be
\label{phi_equiv}
\phi_{\gamma(g)(a),\gamma(g)(a)}(\id_{\gamma(a)})\circ \gamma_{a S(a)}(g)=(-1)^{\epsilon(g)(\epsilon(\kappa)+1)}\phi_{aa}(\id_a)~~,
\ee 
which can be viewed as an equivariance condition for $\phi$. Hence $(S,\phi)$ is a rather special type of equivariant Serre functor 
on $\cT$. A $G$-equivariant cyclic category can be viewed as a $G$-category endowed with such an equivariant Serre functor. 
A general discussion of equivariant Serre functors can be found in Appendix \ref{sec:Serre}. 

\

The following results shows that graded open 2d TFTs with shifts are equivalent with equivariant cyclic categories. This 
translates the physics formalism into a subset of the theory of equivariant Serre functors. 

\paragraph{Proposition} The following data are equivalent: 

(a) A graded open 2d TFT with shifts $(\cG,G,\gamma,s,\tr,\kappa,\epsilon)$

(b) A $G$-equivariant cyclic category $(\cT, G, \gamma,tr,\kappa,\epsilon)$.

\noindent Moreover, the relation between (a) and (b) is given by
$\cG=\cT^\bullet[G]$ (thus $\cT=\cG^0$) and $\tr_a(u)=tr_a(u)$ for all 
$u\in \Hom_{{\cT}^\bullet[G]}^g(a,a)=\Hom_{\cT}^{g-\kappa}(a,\gamma(\kappa)(a))$.

\

\noindent {\em Proof.} 

\

\noindent (1) Given the data at (a), set $\cT=\cG^0$ and define maps
$tr_a:\Hom_{\cT}(a,\gamma(\kappa)(a))\rightarrow \C$ via: \be
\label{Se}
tr_a(u):=\tr_a(\Phi^{-1}(u))~~\forall u\in
\Hom_{\cT}(a,\gamma(\kappa)(a))=\Hom_{\cT^\bullet[G]}^\kappa(a,a)~~,
\ee 
where $\Phi:\cG\stackrel{\sim}\rightarrow \cT^\bullet[G]$ is the
isomorphism of graded categories with shifts constructed in the proof
of the last proposition of Subsection \ref{sec:relation}.

Given $u\in \Hom_{\cT}(a,b)=\Hom_{\cT^\bullet[G]}^0(a,b)$ and
$v\in \Hom_{\cT}(b,\gamma(\kappa)(a))=\Hom_{\cT^\bullet[G]}^\kappa(b,a)$, we
have: \bea
tr_{a}(v\circ u)=\tr_a(\Phi^{-1}(v*u))=\tr_a(\Phi^{-1}(v)\Phi^{-1}(u))=\tr_b(\Phi^{-1}(u)\Phi^{-1}(v))\nn\\
=\tr_b(\Phi^{-1}(u*v))=\tr_b(\Phi^{-1}(\gamma(\kappa)(u)\circ v))=
tr_b(\gamma(\kappa)(u)\circ v)~~,\nn \eea where we used $\epsilon(\deg
u)=\epsilon(0)=0$. This shows that (\ref{tr_cycl}) is
satisfied. Equation (\ref{tr_equiv}) follows trivially from
(\ref{Equiv}). It is also clear that $tr_a$ are nondegenerate.

\

\noindent (2) Given the data at (b), set $\cG:=\cT^\bullet[G]$ and
define maps $\tr_a:\Hom_{\cG}^\kappa(a,a)\rightarrow \C$ via: \be
\label{Sinv}
\tr_a(u):=tr_a(u)~~\forall u\in \Hom_{\cG}^\kappa(a,a)=\Hom_\cT(a,\gamma(\kappa)(a))~~,  \ee 
with extension by zero to other degrees. It is clear that $\tr_a$ are nondegenerate and satisfy the selection rule
(\ref{tr_sel}) as well as
(\ref{Equiv}). To show (\ref{tr_sym}), pick $u\in
\Hom^g_{\cG}(a,b)=\Hom_{\cT}(a,\gamma(g)(b))$ and $v\in
\Hom^h_{\cG}(b,a)=\Hom_{\cT}(b,\gamma(h)(a))$ with $g+h=\kappa$. Then

\

\noindent $ \tr_a(v*u)=tr_a(\gamma(g)(v)\circ
u)=\tr_{\gamma(g)(b)}(\gamma(\kappa)(u)\circ
\gamma(g)(v))=tr_{\gamma(g)(b)}(\gamma(g+h)(u)\circ \gamma(g)(v))=
tr_{\gamma(g)(b)}(\gamma(g)(\gamma(h)(u)\circ v))=
(-1)^{\epsilon(g)(\epsilon(\kappa)+1)} tr_b (\gamma(h)(u)\circ
v)=(-1)^{\epsilon(g)\epsilon(h)} \tr_b (u*v)~~, $

\

\noindent where we used (\ref{tr_cycl}) and (\ref{tr_equiv}) and noticed
that $ (-1)^{\epsilon(g)(\epsilon(\kappa)+1)}=(-1)^{\epsilon(g)\epsilon(h)}$ since $g+h=\kappa$ implies 
$\epsilon(\kappa)=\epsilon(g)+\epsilon(h)$.

\paragraph{Trivial examples} Consider a cyclic group $G=\Z_\omega:=\Z/\omega\Z$ (notice that we allow for the case $s=0$, 
in which case $G=\Z$) and let $[1]:=\gamma(1)$.  Setting
$[n]:=\gamma(n)=[1]^n$, we have $s(n)=s(1)^n$ where 
we use an obvious abuse of notation. The positive integer $\omega$ is
the order of the automorphism $[1]$ (we set $\ord [1]=0$ if $[1]^n\neq
\id_\cG$ for all $n$). Thus a $\Z_\omega$-graded category with shifts is
simply a $\Z_\omega$-graded category endowed with an automorphism $[1]$ of
order $\omega$ and a degree $-1$ isomorphism of functors
$s(1):\id_\cG\stackrel{\sim}{\rightarrow} [1]$.  The later induces linear
isomorphisms $\Hom_\cG^k(a,b[1])\approx \Hom_{\cG}^{k+1}(a,b)$ for all
objects $a,b$. The element $\kappa\in \Z_\omega$ is the common degree of
the traces $\tr_a$.  On the other hand, a $\Z_\omega$-cyclic category is
simply a category $\cT$ endowed with an automorphism $[1]$ of order
$\omega$ and such that $([\kappa],\phi)$ is a Serre functor, where the maps
$\phi_{ab}:\Hom_\cT(a,b)\rightarrow \Hom_{\cT}(b,a[\kappa])^{\rm v}$
correspond to cyclic nondegenerate traces subject to the equivariance
condition
$tr_{a[1]}{u[1]}=(-1)^{\epsilon(1)(\epsilon(\kappa)+1)}tr_a(u)$. Thus
$\cT$ is a Calabi-Yau category of dimension $\kappa$, whose shift
functor has order $\omega$ and whose traces are equivariant in the sense
above (this refines results of \cite{Bocklandt}).  The proposition
above shows that such categories are the same thing as $\Z_\omega$-graded
open 2d TFTs with shifts, the relation in this case being given by
taking the orbit category of $\cT$ with respect to the shift functor $[1]$. 
Notice that the correct grading group of the TFT can be recovered as the
subgroup of $\Aut(\cT)$ generated by $[1]$.  Two
familiar cases are $s=2$ and $s=0$, with morphism $\epsilon$ given by
$\id_{\Z_2}$ and {\rm mod 2} reduction respectively. The first case
corresponds to a generic open 2d TFT, while the second describes
$\Z$-graded theories, such as those associated with the bounded
derived category of coherent sheaves on a generic compact Calabi-Yau
manifold. This last case was discussed in Appendix A of \cite{gen}.

\section{$G$-graded dG categories}

In open topological field theory, $\Z$-graded associative categories
often appear as total cohomology categories of differential graded
(dG) categories. In this section, we extend this construction to
the $G$-graded case. After a brief general discussion, we construct 
a dG category with shifts whose objects are graded integrable modules 
`with extended grading' over a curved differential graded algebra. A particular case of this construction 
will allow us to give a description of graded D-branes in Landau-Ginzburg models 
which manifestly displays the grading by characters of the surviving group 
of vector and axial R-symmetries. 

Let $G$ be an Abelian group endowed with a pairing (=group morphism)
$\cdot:G\times G\rightarrow \Z_2$ and fix an element $\delta\in G$.

\paragraph{Definition} A graded dG category of type $(G,\delta)$ is a $G$-graded $\C$-category $\cD$ 
whose morphism spaces are endowed 
with linear maps $d_{ab}:\Hom_{\cD}(a,b)\rightarrow \Hom_{\cD}(a,b)$ which are homogeneous of degree $\delta$ and
satisfy the following conditions:

(1) $d_{ab}^2=0$ for all $a,b\in \Ob\cD$

(2) $d_{ac}(vu)=d_{bc}(v) u+(-1)^{\delta\cdot \deg v }v d_{ab}(u)$ for
    all homogeneous $u\in \Hom_{\cD}(a,b)$ and $v\in \Hom_{\cD}(b,c)$.

\

\noindent Notice that the definition makes use of the pairing $\cdot:G\times G\rightarrow \Z_2$. In the application to 
open 2d TFTs, this morphism will have the form:
\be
\label{pairing}
g_1\cdot g_2:=\epsilon(g_1)\epsilon(g_2)~~,
\ee
where $\epsilon:G\rightarrow \Z_2$ is the group morphism used in Section 1 and 
in the right hand side we use the ring multiplication of $\Z_2$.

We define the total cohomology category $H(\cD)$ to be the $G$-graded
category having the same objects as $\cD$ and morphism spaces given by
$\Hom_{H(\cD)}(a,b):=\oplus_{g\in G}\Hom_{H(\cD)}^g(a,b)$, where: \be
\nn \Hom_{H(\cD)}^g(a,b):=[\ker(d_{ab})\cap
\Hom_{\cD}^g(a,b)]/[d_{ab}(\Hom_{\cD}^{g-\delta}(a,b))]~~.  \ee
Given any closed morphism $\phi\in \Hom_\cD(a,b)$, we let $\phi^H \in \Hom_{H(\cD)}(a,b)$ be the induced morphism 
in $H(\cD)$.  We also let $Z(\cD)$ be the category obtained from $\cD$ by keeping only closed morphisms. 

A {\em twisted dG automorphism} of $\cD$ is a pair $(F,\chi)$ such that $\chi$ is a non-vanishing complex number and 
$F$ is an automorphism of the underlying associative category, which satisfies the condition: 
\be
F_{ab}(d_{ab}(u))=\chi d_{F(a)F(b)}(F_{ab}(u))
\ee
for all morphisms $u\in \Hom_\cD(a,b)$. Twisted dG automorphisms form a group with the
composition rule $(F,\chi)\circ (F',\chi')=(F\circ F',\chi\chi')$ (the
identity element is $(\id_\cD,1)$). We will denote this group by
$\Aut_{\rm tw}(\cD)$. Notice that any twisted dG automorphism $F$ of $\cD$
induces an automorphism $F^H$ of $H(\cD)$ by passage to cohomology.

\paragraph{Definition} A {\em dG category with shifts} is a system $(\cD,G,\delta, \gamma,s)$ such that:

(1) $\cD$ is a graded dG category of type $(G,\delta)$

(2) $\gamma:G\rightarrow \Aut_{\rm tw}(\cD)$ is a group monomorphism.

(3) $s(g):\id_\cD\stackrel{\sim}{\rightarrow} \gamma(g)$ are {\em
  closed} isomorphisms of functors subject to conditions (\ref{sdeg})
  and (\ref{sdef}) (with $\cG$ replaced with $\cD$).

\

\noindent The closure condition on $s(g)$ means
$d_{a,\gamma(g)(a)}(s_a(g))=0$ for all $a\in \Ob\cD$ and all $g\in
G$. This implies\footnote{It is easy to check that the inverse of any
closed and homogeneous isomorphism $\phi$ is closed and homogeneous of
opposite degree, and induces the inverse of $\phi^H$ on cohomology.}
that $s_a(g)$ descend to isomorphisms $s_a^H(g)$ in $H(\cD)$.

It is easy to see that the total cohomology category $H(\cD)$ of a
graded dG category with shifts is a graded category with shifts, whose
shift functors and suspensions $\gamma^H, s^H$ are induced from $\gamma, s$ by
passage to cohomology. More precisely, $(H(\cD),G,\gamma^H,s^H)$ is a G-graded 
category with shifts.

\

\subsection{Integrable modules over a curved differential graded algebra}

\label{sec:integrable}

In this subsection, we discuss a class of graded dG categories which
generalizes a construction considered in \cite{Block}. As we will see below, this
affords a `manifestly symmetric' description of the dG
category of graded D-branes in Landau-Ginzburg models.

\paragraph{Modules with extended grading over a graded algebra}

Let $G_V$ be an Abelian group and $B$ a $G_V$-graded associative algebra. Given another Abelian group $G$ and a morphism 
$\psi:G_V\rightarrow G$, a right {\em extended-graded module} of type $(G,\psi)$ over $B$ is a right $B$-module $M$
endowed with a vector space decomposition: 
\be
\label{M_decomp}
M=\oplus_{g\in G}{M^g}~~,
\ee
such that $M^gB^k\subset M^{g+\psi(k)}$ for all $g\in G$ and $k\in G_V$. Given such a module, we set $\deg(m)=g$ for $m\in M^g$. 
A decomposition (\ref{M_decomp}) with this property will be called a {\em extended graded structure} of type $(G,\psi)$ on $M$.
Given two such modules $M,N$, we define: 
\be
\nn
\Hom_B^g(M,N):=\{f\in \Hom_B(M,N)|f(M^h)\subset M^{g+h}~~\forall h\in G\}
\ee
and $\Hom_B^{\gr} (M,N):=\oplus_{g\in G} \Hom_B^g(M,N)$. This gives a $G$-graded category denoted $\Gr^{G,\psi}_B$.
Notice that $\Gr^{G_V,\id_{G_V}}_B$ coincides with the usual category of graded $B$-modules.

\paragraph{Twist functors and suspensions.}
  
Given $g\in G$, we let $(g)\in \Aut^0[\Gr^{G,\psi}_B] $ denote the {\em twist functor by $(g)$}. This acts on objects
via $M(g)^h:=M^{h+g}$ and takes a morphism $u\in \Hom_B^h(M,N)$ into the same 
map $u$ but viewed as an element of $\Hom_B^h(M(g),N(g))$.  We have $(g)\circ (h)=(g+h)$ for all $g,h\in G$, as well as 
the relation $\Hom^h_B(M,N(g))=\Hom^{h+g}_B(M,N)$.   

Let $s_M(g)$ be the $g$-th {\em suspension morphism} of $M$, i.e. the identity morphism of $a$ viewed as an element of $\Hom^{-g}_B(M,M(g))$. 
These maps define isomorphisms of functors $s(g):\id_{\Gr^{G,\psi}_B}\stackrel{\sim}{\rightarrow} (g)$.  
Setting ${\tilde \gamma}(g):=(g)$, it is clear that (\ref{sdef}) holds. 
Thus $(\Gr^{G,\psi}_B, G, {\tilde \gamma},s)$ is a graded category with shifts. 
We let $\gr_B^{G,\psi}$ denote the full subcategory of $\Gr_B^{G,\psi}$ consisting of finitely-generated modules. 
This is again a graded category with shifts.

\paragraph{Free extended-graded modules} An extended-graded module $M\in \Ob[\Gr_B^{(G,\psi)}]$ will be called {\em free} 
if it admits a basis $e_s$ all of whose elements are $G$-homogeneous. In this case, we have 
a decomposition: 
\be
\label{Mbasis}
M=\oplus_{s}{e_s B}
\ee
and 
\be
\label{Mfree}
M^g=\oplus_{s,k:g_s+\psi(k)=g} e_s B^k~~.
\ee

\paragraph{Change of grading.}

The collection of pairs $(G,\psi)$ where $G$ is a group and $\psi:G_V\rightarrow G$ is a group morphism forms a category $\Lambda$ 
whose morphisms $(G,\psi)\rightarrow (G',\psi')$ are maps $\mu:G\rightarrow G'$ such that $\mu\circ \psi=\psi'$. Given 
$\mu\in \Hom_\Lambda((G,\psi),(G',\psi'))$ and $M$ in $\Gr^{G,\psi}_B$, we let $\mu_*(M)$ be the object of $\Gr^{G',\psi'}_B$
defined by the decomposition $\mu_*(M)=\oplus_{g'\in G'}{\mu_*(M)^{g'}}$, where 
$\mu_*(M)^{g'}:=\oplus_{g\in G:\mu(g)=g'}M^g$ (it is easy to check that the 
condition $\mu^*(M)^{g'}B^k\subset \mu^*(M)^{g'+\psi'(k)}$ is satisfied). 
Notice that we set $\mu_*(M)^{g'}=0$ if $g'\notin \mu(G)$. 
It is clear that a homogeneous 
morphism $u\in \Hom^h_{\Gr^{G,\psi}_B}(M,N)$ satisfies $u(\mu_*(M)^{g'})\subset \mu_*(M)^{g'+\mu(h)}$, 
and we let  $\mu_*(u)\in \Hom^{\mu(h)}_{\Gr^{G',\psi'}_B}(\mu_*(M),\mu_*(N))$ be the morphism of $\Gr^{G',\psi'}_B$
obtained in this manner. Then  $\mu_*$ gives a functor from $\Gr^{G,\psi}_B$  to $\Gr^{G',\psi'}_B$. Given composable morphisms 
$\mu,\nu$ in $\Lambda$, one has $(\mu\circ \nu)_*=\mu_*\circ \nu_*$. Further, the 
identity automorphism of $(G,\psi)$ induces the identity automorphism of $\Gr^{G,\psi}_B$. These observations imply that  any isomorphism 
$\mu:(G,\psi)\rightarrow (G',\psi')$ of $\Lambda$ induces an isomorphism of categories $\mu_*:\Gr^{G,\psi}_B\rightarrow \Gr^{G',\psi'}_B$, so 
we can identify the categories of graded modules defined by isomorphic pairs $(G,\psi)$.

\paragraph{Extensions of grading.}

Consider a $G_V$-graded algebra $B$, and let $B^\hash$ be
the associative algebra obtained from $B$ by forgetting the grading.
Fixing an Abelian group $G_A$, consider the category
$\Gr^{G_A}_{B^\hash}$ of $G_A$-graded modules over $B^\hash$ (where
$B^\hash$ is viewed as a graded algebra concentrated in degree zero). 
The objects of $\Gr^{G_A}_{B^\hash}$ are $B^\hash$-modules $M$ endowed with a decomposition of the form 
$M=\oplus_{\alpha\in G_A}M_\alpha$ into $B^\hash$-{\em submodules}.

Consider a short exact sequence: 
\be
\label{extension}
(E)~~0\rightarrow G_V\stackrel{\psi}{\rightarrow} G
\stackrel{\epsilon}{\rightarrow} G_A\rightarrow 0~~
\ee 
representing an extension class $\xi\in \Ext_\Z^1(G_A, G_V)$.  Notice
that $\Gr^{G_A}_{B^\hash}=\Gr_B^{(G_A,0)}$, where $0$ is the null
morphism from $G_V$ to $G_A$.  Since $\epsilon\circ \mu=0$, we can
view $\epsilon$ as a morphism in $\Hom_\Lambda((G,\psi),(G_A,0))$.
This induces a functor $\epsilon_*:\Gr_B^{(G,\psi)}\rightarrow
\Gr^{G_A}_{B^\hash}$. Since $\epsilon$ is surjective, we have
$\epsilon_*(M)=M$ as vector spaces, and we set
$M_\alpha:=\epsilon_*(M)^\alpha=\oplus_{g\in G:\epsilon(g)=\alpha}
M^g$ for all $\alpha\in G_A$.

Given an object $M$ of $\Gr^{G_A}_{B^\hash}$, an {\em extension of
grading } of $M$ along the exact sequence $(E)$ is an extended graded
module structure $M=\oplus_{g\in G}M^g$ of type $(G,\psi)$ such that
$M_\alpha=\oplus_{\epsilon(g)=\alpha}M^g$ for all $\alpha\in G_A$,
i.e. such that the $\epsilon_*$-image of the extended-graded $B$-module
obtained in this manner recovers $M$. Such a structure exists iff $M$
belongs to the image of the functor
$\epsilon_*:\Gr^{(G,\psi)}_B\rightarrow \Gr^{G_A}_{B^\hash}$, in which
case we say that $M$ is {\em gradable along (E)}.

Recall that extensions of $G_A$ by $G_V$ form a category. Given
another extension $(E')~~0\rightarrow G_V\stackrel{\psi'}{\rightarrow}
G' \stackrel{\epsilon'}{\rightarrow} G_A\rightarrow 0$, the morphisms
$(E)\rightarrow (E')$ are given by maps $\mu:G\rightarrow G'$ such
that $\mu\circ \psi=\psi'$ and $\epsilon'\circ \mu=\epsilon$. In
particular, $\mu$ is a morphism in $\Lambda$ from $(G,\psi)$ to
$(G',\psi')$, so it induces a functor $\mu_*:\Gr_B^{G,\psi}\rightarrow
\Gr_B^{G',\psi'}$. We thus have a triple of functors
$(\mu_*,\epsilon_*,\epsilon'_*)$ which satisfy $\epsilon'_*\circ
\mu_*=\epsilon_*$, and $\mu_*$ is an isomorphism of categories when
$\mu$ is an isomorphism of groups.  When $\mu$ is an isomorphism, this
shows that $\epsilon_*$ and $\epsilon'_*$ can be identified
consistently with the identification of $\Gr_B^{G,\psi}$ and $\Gr_B^{G',\psi'}$
via $\mu_*$. It follows that extensions of grading are essentially
determined by the extension class $\xi$ of the sequence $(E)$.

\paragraph{Component description.}

Given $M\in \Ob [\Gr^{G_A}_{B^\hash}]$ endowed with an extension of grading along $(E)$, the 
$G_A$-components $M_\alpha$ become {\em graded} $B$-modules via
the following construction. Let us pick elements $g_\alpha\in G$ such
that $\epsilon(g_\alpha)=\alpha$ for each $\alpha\in G_A$. Then the
$\epsilon$-preimage of $\alpha$ consists of the elements
$g=g_\alpha+\psi(k)$ with $k\in G_V$. Defining: \be \label{Mdec}
M_\alpha^k:=M^{g_\alpha+\psi(k)}~~\forall \alpha\in G_A~,\forall k\in
G_V~~, \ee it is clear that $M_\alpha^kB^l\subset
M_\alpha^{k+l}$. Thus the decomposition $M_\alpha =\oplus_{g\in G:\epsilon(g)=\alpha}M^g=\oplus_{k\in G_V}M_\alpha^k$ gives a
$G_V$-grading of the module $M_\alpha$, and $M$ can be viewed as direct
sum of graded $B$-modules. Given another choice of elements $g'_\alpha\in G$ 
such that $\epsilon(g'_\alpha)=\alpha$, we have $g'_\alpha-g_\alpha=\psi(k_\alpha)$ for 
some $k_\alpha\in G_V$, so the $B$-graded module structures $(M')_\alpha^k=M^{g'_\alpha+\psi(k)}$ determined 
by this new choice on $M_\alpha$ are related to (\ref{Mdec}) via the twists:
\be
\label{ktf}
M'_\alpha =M_\alpha(k_\alpha)~~.
\ee
It follows that the $G_V$-gradings on $M_\alpha$ are determined by the $G$-grading on $M$ up to independent twists. 
We stress that the gradings on $M_\alpha$ depend on the choice of $g_\alpha$.

Conversely, any collection $(M_\alpha)_{\alpha\in G_A}$ of $G_V$-graded modules 
defines an object $M:=\oplus_{\alpha\in G_A}M_\alpha$ of $\Gr^{G_A}_{B^\hash}$ 
which is $(E)$-gradable for any fixed extension (\ref{extension}). 
To see this, pick elements $g_\alpha\in G$ such that $\epsilon(g_\alpha)=\alpha$.
Given $g\in G$, set $\alpha:=\epsilon(g)$. Then $g-g_{\alpha}\in \ker \epsilon=\im \psi$, so there exists a unique $k\in G_V$ such that $g=g_\alpha+\psi(k)$.
This shows that the map $(k,\alpha)\in G_V\times G_A\stackrel{f}{\rightarrow} \psi(k)+g_\alpha \in G$ is a bijection.  
Defining $M^g:=M_{\alpha}^{k}$ where $(k,\alpha)=f^{-1}(g)$, we can thus 
write the linear subspace decomposition $M=\oplus_{\alpha\in G_A, k\in G_V}M_\alpha^k$ as $M=\oplus_{g\in G} M^g$. It is clear that the 
condition $M_\alpha^kB^l\subset M_\alpha^{k+l}$ implies $M^gB^l\subset M^{g+\psi(l)}$. Thus $M$ endowed with this $G$-grading 
is an extended graded module of type $(G,\psi)$ over $B$, and 
$\epsilon_*(M)$ recovers the original object of $\Gr^{G_A}_{B^\hash}$.
If we pick other elements $g'_\alpha\in \epsilon^{-1}(\{\alpha\})$, then $g'_\alpha-g_\alpha=\psi(k_\alpha)$ 
for some $k_\alpha\in G_V$ and we find that the $G$-grading determined by $g'_\alpha$ is related to that given by $g_\alpha$ 
via: 
\be
\label{gtf}
M'^g=M^{g-\psi(k_{\epsilon(g)})}~~.
\ee
Hence the choice of the extension $(E)$ determines the $G$-grading on $M$ up such transformations. This discussion 
establishes the following:

\paragraph{Proposition} The following statements are equivalent for any object $M$ of $\Gr^{G_A}_{B^\hash}$:

(a) $M$ is gradable along  some extension $(E)$ of type (\ref{extension}) 

(b) $M$ is gradable along  any extension $(E)$ of type (\ref{extension})

(c) Each of the $B^\hash$-submodules $M_\alpha$ admits a structure of graded $B$-module.

\noindent In this case, the extension $(E)$ together with the graded $B$-module structures on $M_\alpha$ determine the $G$-grading on $M$ up to 
transformations of the form (\ref{gtf}), while  $(E)$ and the $G$-grading on $M$ determines the $G_V$-gradings on $M_\alpha$ up to 
independent twists of the form (\ref{ktf}). 

\

\noindent The proposition shows that the notion of module with extended grading is essentially the same as a collection of graded $B$-modules. 
The gain afforded by the former language is that it manifestly displays the larger group $G$. 

\paragraph{Component description of free extended-graded modules}

Consider an extended-graded module of type $(G,\psi)$ over $B$ and fix an extension $(E)$ as in (\ref{extension}). Also pick elements $g_\alpha\in G$ such 
that $\epsilon(g_\alpha)=\alpha$ for all $\alpha\in G_A$. Then it is easy to see that $M$ is free as an extended-module iff all of 
the graded $B$-modules $M_\alpha$ determined by $g_\alpha$ are free. We leave the details as an exercise for the reader. 

\paragraph{Curved differential graded algebras.}

Consider an Abelian group $G_V$ endowed with a pairing $\cdot:G_V\times
G_V\rightarrow \Z_2$. A  $G_V$-graded {\em curved differential graded algebra} (CdGA) over $\C$ is a
triple $(B,d,c)$ where $B$ is a $G_V$-graded associative algebra over
$\C$, $c$ is a homogeneous element of $B$ and $d$ is a homogeneous
derivation of $B$ such that $d^2(b)=[c,b]$ for all $b\in B$, where
$[b_1,b_2]:=b_1b_2-(-1)^{\deg b_1\cdot \deg b_2}b_2b_1$ denotes the graded commutator  (computed using the pairing on $G_V$). 
If $c\neq 0$, then we let $h\in G_V$ denote the degree of $c$.  The Leibniz condition for the derivation $d$ takes the
form:
\be
\label{graded_Leibniz}
d(b_1b_1)=d(b_1)b_2+(-1)^{\deg(d)\cdot \deg(b_1)}b_1d(b_2)~~.
\ee
Notice that we must have $2\deg(d)=h$ unless $c=0$ or $d$ is identically zero. 
When $c=0$, the pair $(B,d)$ is a $G_V$-graded differential algebra (a $G_V$-graded differential category with one object). 
The case $d=0$ is also special. 
In this case, $(B,0,c)$ is a curved dGA iff $c$ is a homogeneous 
{\em central} element of $B$, since the condition $d^2(b)=[c,b]$ becomes $[c,b]=0$ for all $b\in B$. Then (\ref{graded_Leibniz}) is trivially 
satisfied and there are no constraints on the degree of $c$.

\paragraph{Integrable modules over a curved dGA.}

Let us fix an Abelian group $G$ with a pairing
$\cdot:G\times G\rightarrow \Z_2$ and a group morphism
$\psi:G_V\rightarrow G$. Also fix an element $\delta\in G$. 
A {\em connection of degree $\delta$ } on $M\in \Ob [\gr^{G,\psi}_B]$    
is a $\C$-linear map $D:M\rightarrow M$ satisfying the homogeneity condition:
\be
\deg D (m)=\deg m+\delta~~
\ee
and the Leibniz identity: 
\be
\label{Leib}
D(mb)=D(m)b+(-1)^{\delta\cdot \deg m} ~m d(b)~~.
\ee
These relations require $\delta=\psi(\deg d)$ unless $d=0$, in which case $\delta$ is unconstrained. 

An {\em integrable module of type $(G,\psi,\delta)$} 
over $B$ is a pair $\M:= (M,D)$ such that $M$ is a finitely-generated right extended-graded $B$-module of type $(G,\psi)$ 
and $D$ is a connection of degree $\delta$ on $M$, subject to the condition:
\be
\label{int}
D^2(m)=mc~~\forall m\in M~~.
\ee 
This equation requires $2\delta=\psi(h)$, which is compatible with the previous constraint 
$\delta=\psi(\deg d)$ when $d\neq 0$ (since $2\deg d=h$ in that case). Notice that (\ref{int}) constrains $\delta$ also for the case $d=0$. 

It is easy to check that integrable modules of type $(G,\psi,\delta)$
form a dG category $\DG_c^{G,\psi,\delta}(B)$ of type $(G,\delta)$, with morphism spaces given by
$\Hom_{\DG_c^{G,\psi,\delta}(B)}(\M,\N):=\Hom_B^\gr(M,N)$ and differentials defined through
$d_{\M\N}(u):=D_N\circ u -(-1)^{\delta\cdot \deg u} u\circ D_M$ for
homogeneous $u\in \Hom_B^\gr (M,N)$.

\paragraph{A particular case} 

A particularly simple case arises when the differential on $B$
vanishes (thus $c$ must be a homogeneous central element of $B$, and we assume $c\neq 0$). In this case,
$D$ is a homogeneous module endomorphism of $M$ of degree $\delta$ which
satisfies (\ref{int}); this requires $2\delta=\psi(h)$ where $h=\deg c$. 

The twist functors of $\gr^{G,\psi}_B$ induce twist functors on
$\DG_c^{G,\psi,\delta}(B)$ as follows. For any object $\M=(M,D)$, we
set $\M(g):=(M(g),D(g))$, where $D(g)$ is the result of acting on the
module endomorphism $D$ with the twist functor of $\gr^{G,\psi}_B$. We
let $(g)$ act on the morphisms of $\DG_c^{G,\psi,\delta}(B)$ in the
same way as in $\Gr^{G,\psi}_B$.  It is easy to check that $(g)$ are
dG-automorphisms, making $\DG_c^{G,\psi,\delta}(B)$ into a dG
category of type $(G,\delta)$.

The maps $s_\M(g):=s_M(g)\in \Hom_B^{-g}(M,M(g))=\Hom_{\DG_c^{G,\psi,\delta}(B)}^{-g}(\M,\M(g))$ 
give isomorphisms of functors from $\id_{\DG_c^{G,\psi,\delta}(B)}$ to $(g)$. In general, these isomorphisms are 
not closed, so they cannot be used to make $\DG_c^{G,\psi,\delta}(B)$ into a graded category with shifts. 
Indeed, they satisfy $D_{M(g)}\circ s_M(g)=s_M(g)\circ D_M$ rather than the closure condition 
$d_{\M\M(g)}(s_M(g))=D_{M(g)}\circ s_M(g)-(-1)^{\delta\cdot g}s_M(g)\circ D_M=0$. This can be remedied in the following 
situation, which will be relevant below. 

Let us assume given an exact sequence of the form (\ref{extension}) with $G_A=\Z_2$ and 
consider the pairing on $G$ given in (\ref{pairing}). Define automorphisms $\gamma(g)$ of the underlying associative category of 
$\DG_c^{G,\psi,\delta}(B)$ via: 
\be
\label{gamma_def}
\gamma(g)(M,D)=(M(g),(-1)^{\epsilon(g)\epsilon(\delta)}D(g))~~
\ee
and:
\be
\gamma_{\M\N}(g)(u):=u(g)~~\forall u\in \Hom_{\DG_c^{G,\psi,\delta}(B)}(\M,\N).
\ee
We have $\gamma(g_1)\gamma(g_2)=\gamma(g_1+g_2)$ for all $g_1,g_2\in G$. 
Then $\gamma(g)$ are twisted dG functors, namely they satisfy: 
\be
d_{\gamma(g)(\M)\gamma(g)(\N)}\circ \gamma_{\M\N}(g)=\chi(g)\gamma_{\M\N}(g)\circ d_{\M\N}~~
\ee
for the character $\chi(g)=(-1)^{\epsilon(g)\epsilon(\delta)}$ of $G$. On the other hand, 
the maps $s_\M(g)=s_M(g)$ {\em are} closed when viewed as elements of 
$\Hom^{-g}_{\DG_c^{G,\psi,\delta}(B)}(\M,\gamma(g)(\M))$, and we find that $(\DG_c^{G,\psi,\delta}(B),G,\delta,\gamma,s)$ is a 
dG category with shifts.  Passing to the total 
cohomology category $\cG:=H(\DG_c^{G,\psi,\delta}(B))$, we have induced automorphisms 
$\gamma^\cG(g)$ of $\cG$ and morphisms of functors $s^\cG(g):\id_\cG\rightarrow \gamma^\cG(g)$ of degree $-g$ making  
$(\cG,G,\gamma^\cG, s^\cG)$ into a graded category with shifts. 

Consider the dG category $\DG_c^{\Z_2,0,\epsilon(\delta)}(B)$, 
where $0$ is the null morphism form $G_V$ to $G_A=\Z_2$. 
This consists of pairs $\M=(M,D)$ with $M\in \Ob[ \gr_{B^\hash}^{\Z_2}]$ and $D\in \End_{B^\hash}(M)$ 
a homogeneous module endomorphism of degree $\epsilon(\delta)$ satisfying $D^2=c$ (recall that $B^\hash$ is concentrated in degree zero).
The homogeneity condition on $D$ means $D(M_\alpha)=M_{\alpha+\epsilon(\delta)}$. 
The functor $\epsilon_*:\Gr_B^{G,\psi}\rightarrow \Gr_{B^\hash}^{\Z_2}$ 
induces a dG functor ${\tilde \epsilon}_*:\DG_c^{G,\psi,\delta}(B) \rightarrow \DG_c^{\Z_2,0,\epsilon(\delta)}(B)$, 
defined through ${\tilde \epsilon}_*(M,D)=(\epsilon_*(M),\epsilon_*(D))$ on objects and 
${\tilde \epsilon}_*(u):=\epsilon_*(u)$ on morphisms.  An object of $\DG_c^{\Z_2,0,\epsilon(\delta)}(B)$ is 
called gradable along $(E)$ if it lies in the image of this functor. 
Picking elements $g_\alpha\in G$ such that $\epsilon(g_\alpha)=\alpha$, consider the $G_V$-grading of $M_\alpha$ 
defined as in the previous subsection.  Then it is easy to see that gradability of $\M$ reduces 
to gradability of $M$ (i.e. the condition that $M$ are graded $B$-modules) 
and the condition that the components $D_{\alpha, \alpha+\epsilon(\delta)}\in \Hom_{B}(M_\alpha,M_{\alpha+\epsilon(\delta)})$ 
of $D$ are $G_V$-homogeneous of degree $\delta+g_\alpha-g_{\alpha+\epsilon(\delta)}$.

\section{Application to graded B-type branes in Landau-Ginzburg models}
\label{sec:LG}

In this section, we reconsider the case of graded B-type branes in
Landau-Ginzburg models, giving a complete treatment of the vector and
axial gradings.  While this was already studied in \cite{Walcher,
Orlov3, saito} from various perspectives, we will see in a moment that
the grading induced by vector and axial R-symmetries on the
corresponding open 2d TFT is a bit subtle. The reason is that the total grading arises from a
group extension of the axial $\Z_2$-grading by the vector
$\Z$-grading, and this extension is nontrivial when the degree of the
bulk superpotential is odd. We will also give a complete description
of the open 2d TFT defined by graded D-branes, as well as its precise
relation with the triangulated category of matrix factorizations
considered in \cite{Orlov3}. Using this correspondence will allow us
to extract a precise proposal for the Serre functor on the latter.

Consider the B-twisted topological Landau-Ginzburg model with target
$\C^n$ and coordinate ring $B:=\C[\phi_1\ldots \phi_n]$. The $U(1)$
vector R-symmetry induces a $G_V=U(1)^*=\Z$-grading of $B$ with $\deg
\phi_i=q_i$, where we take $q_i$ to be integral (see Appendix \ref{sec:worldsheet} for the worldsheet realization). 
From the worldsheet perspective, this corresponds to a  normalization in which the bulk
superpotential $W\in B$ has integral degree $h$. We endow the group $G_V=\Z$ with the trivial 
pairing $m\cdot n=0$ (this is insures that the graded commutator in $B$ is the usual commutator). 
Then the bulk model is encoded by the commutative $\Z$-graded curved dGA $(B,0,W)$, 
which has trivial differential.

\subsection{The dG category of graded D-branes} 

As explained in Appendix \ref{sec:worldsheet}, the axial R-symmetry is
broken by the bulk superpotential to a subgroup $\Gamma_A\approx \Z_2$,
so the surviving vector-axial R-symmetry group is $\Gamma=\Gamma_V\times
\Gamma_A=U(1)\times \Z_2$.  It turns out that the subgroup $\Gamma_0$
which acts trivially on the boundary data depends on the
parity of $h$. Namely, $\Gamma_0$ is trivial for even $h$, while for
odd $h$ one finds that $\Gamma_0$ is a $\Z_2$ subgroup. As a consequence, the
faithfully acting group $\Gamma'_h:=\Gamma/\Gamma_0$ is given by:
\be
\nn
\Gamma'_h=\left\{\begin{array}{lll}
U(1)\times \Z_2& {\rm ~for~} h\in 2\N~, \\
U(1)  &{\rm ~for~} h\in 2\N+1\ 
\end{array}\right.~~, 
\ee
with group of characters:
\be
\label{G_h}
G_h=(\Gamma'_h)^*=\left\{\begin{array}{lll}
\Z\times \Z_2& {\rm ~for~} h\in 2\N~, \\
\Z  &{\rm ~for~} h\in 2\N+1\ 
\end{array}\right.~~.
\ee
When $h$ is odd, the vector and axial gradings on the spaces of boundary and boundary condition changing observables 
combine into a single $\Z$-grading and thus they cannot be treated 
as independent. One has a group extension: 
\be
\label{LG_extension}
0\rightarrow \Z \stackrel{\psi_h}{\rightarrow} G_h\stackrel{\epsilon_h}\rightarrow \Z_2\rightarrow 0~~,
\ee
which is trivial only for even $h$. Here $G_V=\Z$ and $G_A=\Z_2$. The morphisms $\psi_h, \epsilon_h$ are given
by (see Appendix \ref{sec:worldsheet}):
\be
\label{psi_h}
\psi_h(k)=\left\{\begin{array}{lll}
(k,0)& {\rm ~for~} h\in 2\N~, \\
2k  &{\rm ~for~} h\in 2\N+1\ 
\end{array}\right.~~.
\ee
and
\be
\label{epsilon_h}
\left\{\begin{array}{lll}
\epsilon_h(k,\alpha)=\alpha & {\rm ~for~} h\in 2\N, \\
\epsilon_h(k)=k~(\mod~2) &{\rm ~for~} h\in 2\N+1\ 
\end{array}\right.~~. 
\ee
We will use the pairing (\ref{pairing}) on the group $G_h$.

The most general topological D-branes of the model form the the full dG category $\DG_W(B^\hash)$ of $\DG_W^{\Z_2,0,{\hat 1}}(B)$ 
consisting of free graded $B^\hash$-modules (recall that $B^\hash$ is viewed as a graded algebra concentrated in degree zero). 
This is the well-known dG category of matrix factorizations discussed in 
\cite{Orlov1, Orlov2}. 

As shown in Appendix \ref{sec:worldsheet}, `graded
topological D-branes' (i.e. those topological D-branes which preserve
the vector R-symmetry) correspond to objects of  $\DG_W(B^\hash)$ which are freely extended-graded along the extension 
(\ref{LG_extension}). More precisely, such branes are described by integrable $B$-modules $\M=(M,D)$ of type
$(G_h,\psi_h,\delta_h)$, where $M$ is free as an extended-graded module and:
\!\!\!\!\!\!\!\!\be
\label{degD}
\delta_h:=\deg(D)=\left\{\begin{array}{lll}
(\frac{h}{2},{\hat 1})\in \Z\times \Z_2 & {\rm ~for~} h\in 2\N~~, \\
h\in \Z  &{\rm ~for~} h\in 2\N+1\ 
\end{array}\right.~~.
\ee
Notice that $\epsilon_h(\delta_h)={\hat 1}$, and $2\delta_h=\psi_h(h)$, as required. 
Since $(B,0,W)$ has trivial differential, $D$ is a simply a module
endomorphism of $M$ of $G_h$-degree $\delta_h$, which satisfies $D^2=W$.
This gives the following description, which is quite similar in spirit to that used in \cite{CIL2, CIL3, Diac, Block} for 
the case of B-type branes on Calabi-Yau manifolds. We can take it as a mathematical definition: 

\

{\em The dG category of graded D-branes defined by the LG model with bulk data $(B,W)$ is the full dG subcategory 
$\DG_W^\gr(B)$ of $\DG_W^{G_h,\psi_h,\delta_h}(B)$ consisting of free extended-graded modules. }

\

Notice that this description manifestly displays the vector-axial
grading group $G_h$.  The total cohomology category
$\cG:=H(\DG_W^\gr(B))$ is the category of boundary sectors. The dG
category $\DG_W^\gr(B)$ is a dG category with shifts of type
$(G_h,\delta_h, \gamma_h,s_h)$, where the twisted dG automorphisms
$\gamma_h(g)$ are defined as in (\ref{gamma_def}) and the suspensions
$s_h(g)$ are as described in the previous section.  These descend to
automorphisms $\gamma_h^\cG(g)$ of $\cG$ and morphisms of functors
$s_h^\cG(g):\id_{\cG}\rightarrow \gamma^\cG(g)$, making $\cG$ into a
graded category with shifts of type
$(G_h,\gamma_h^\cG,s_h^\cG)$. Its null restriction
$\cT:=\cG^0=H^0(\DG_W^\gr(B))$ is the triangulated category of matrix
factorizations considered in \cite{Orlov3} and \cite{saito}.

\subsection{Shift functor and distinguished twist functor}

Consider the functors $T_h:=\gamma_h(g_T)$, $\tau_h:=\gamma_h(g_\tau)$ and $\rho_h=\gamma_h(g_\rho)$, 
where $g_T$, $g_\tau$ and $g_\rho$ are given by:   
\be
\label{gTtau}
\left\{\begin{array}{lll}
g_\tau=(1,{\hat 0})~~,~~g_T=(\frac{h}{2},{\hat 1})~~,~~g_\rho=(0,{\hat 1}) & {\rm ~for~} h\in 2\N~, \\
g_\tau=2~~~~~,~~~~g_T=h~~~~,~~~~~g_\rho=1 &{\rm ~for~} h\in 2\N+1\ 
\end{array}\right.
\ee
We have $g_T=\delta_h$, $\epsilon_h(g_\tau)={\hat 0}$ and $\epsilon_h(g_T)=\epsilon_h(g_\rho)={\hat 1}$.
Notice the relation $g_T^2=g_\tau^h$, which implies 
\be
\label{Ttau}
T_h^2=(\tau_h)^h~~, 
\ee
as well as the relation $g_T =g_\rho +[\frac{h}{2}]g_\tau$, which gives
\be
\label{rho}
T_h=\rho_h\circ (\tau_h)^{[\frac{h}{2}]}~~.
\ee
Here $[~]$ denotes the integer part. Using these equations, it is easy to check that the subgroup of $\Aut_{\rm  tw}(\DG_W^\gr(B))$ 
generated by $T_h$ and $\tau_h$ is isomorphic with $G_h$. In fact, the representation $\gamma_h$ 
takes the form: 
\be
\label{gamma_LG}
\left\{\begin{array}{lll}
\gamma_h(k,\alpha)= \tau_h^k \rho_h^\alpha =T_h^\alpha\tau_h^{k-\frac{\alpha h}{2}}& {\rm ~for~} h\in 2\N~, \\
\gamma_h(k)=\rho_h^k =T_h^k\tau_h^{-\frac{k(h-1)}{2}}&{\rm ~for~} h\in 2\N+1\ 
\end{array}\right.
\ee

Passing to zeroth cohomology, the functor $T_h$ induces the shift functor $T_h^\cT=[1]$ of the triangulated category $\cT$.
It also turns out that functor $\tau_h^\cT$ induced by $\tau_h$ on $\cT$ coincides with the `distinguished' 
twist functor considered in \cite{Orlov3, saito} (see the last subsection below). 

\subsection{Traces}
Given an object $\M=(M,D)$ of $\DG_W^\gr(B)$, the free graded module $M$ can be decomposed as: 
\be
\nn
M=\oplus_s e_s B~~,
\ee
where $e_s$ is a $G$-homogeneous basis of $M$ with $\deg e_s=g_s\in G$. 
With respect to such a basis, a module endomorphism $u\in \Hom_B(M,M)$ has a matrix with entries $u_{st}\in B$ determined by the decomposition: 
\be
\nn
u(e_s)=\sum_{t} e_t u_{ts}~~.
\ee
This allows one to define a linear map $\Str_\M:\Hom_{\DG_W^\gr(B)}(\M,\M)\rightarrow B$ via: 
\be
\nn
\Str_\M(u):=\sum_{s}(-1)^{\epsilon_h(g_s)}u_{ss}~~.
\ee
When $u$ is homogeneous of degree $g\in G$, we have $u_{ts}=0$ unless $\psi_h(\deg u_{ts})=g+g_s-g_t$. 
This implies $\psi_h(\deg(\Str_\M(u))=g$ unless $\Str_\M(u)= 0$. In particular, we have $\Str_\M(u)=0$ unless $\epsilon(g)={\hat 0}$. 
It is easy to check the relations:
\be
\label{Str_sym}
\Str_\M(uv)=(-1)^{\epsilon_h(\deg u)\epsilon_h(\deg v)} \Str_\N(vu)
\ee
for homogeneous $u\in \Hom_{\DG_W^\gr(B)}(\N,\M)$ and   $v\in \Hom_{\DG_W^\gr(B)}(\M,\N)$, as well as the relation:
\be
\label{Str_twist}
\Str_{\M(g)}(u(g))=(-1)^{\epsilon_h(g)}\Str_\M(u)~~,
\ee
where $(g)$ is the twist by $g$. 

The category $Z(\DG_W^\gr(B))$  is endowed with the traces of \cite{KY, LG2}:
\be
\label{Ktrace}
\tr_\M(u):=\oint{\frac{d\phi_1\ldots d\phi_n}{(2\pi i)^n} \frac{\Str_\M [(\partial D)^n u]}{\partial_1W\ldots \partial_n W}}~~~~~~~~~
(u\in Z(\Hom_{\DG_W^\gr}(\M,\M)))~~~~~,
\ee 
where $\M=(M,D)$ and $(\partial D)^n:=\partial_1 D\ldots \partial_n D$. Notice that $\partial_i D\in \Hom_B^{\delta_h -\psi_h(q_i)}(M,M)$, so 
$\deg (\partial D)^n=n\delta_h -\psi_h(\sum_{i=1}^n{q_i})$. The traces are easily seen to satisfy $\tr_\M (d_{\M\M}(u))=0$ as well as
\be
\label{LG_sym}
\tr_\M(uv)=(-1)^{\epsilon_h(\deg u)\epsilon_h(\deg v)}\tr_\N(vu)
\ee
for homogeneous $u\in \Hom_{\DG_W^\gr(B)}(\N,\M)$ and   $v\in \Hom_{\DG_W^\gr(B)}(\M,\N)$. They also satisfy the selection rule: 
\be
\label{srule}
\tr_\M(u)=0~~{\rm unless}~~\deg(u)=\kappa_h~~, 
\ee
where $\kappa_h\in G_h$ is given by $\kappa_h=n\delta_h-\psi_h(\sum_{i=1}^n q_i)$, i.e.: 
\!\!\!\!\!\!\!\!\be
\label{kappa_LG}
\kappa_h=\left\{\begin{array}{lll}
(\frac{h}{2}{\hat c}, n~\mod ~2)=(\frac{nh}{2}-\sum_{i=1}^n{q_i},~~n~\mod ~2)\in \Z\times \Z_2 & {\rm ~for~} h\in 2\N~, \\
h{\hat c}=nh-2\sum_{i=1}^n{q_i} \in \Z  &{\rm ~for~} h\in 2\N+1\ 
\end{array}\right.~~.
\ee
Here: 
\be
\label{hat_c}
{\hat c}=n-\frac{2}{h}\sum_{i=1}^n{q_i}
\ee
is the conformal anomaly. To establish (\ref{srule}), notice that the integral $\oint{\frac{d\phi_1\ldots d\phi_n}{(2\pi i)^n} {f(\phi_1\ldots \phi_n)}}$ of a homogeneous 
rational function of $\phi_1\ldots \phi_n$ vanishes unless $f$ has degree $-\sum_{i=1}^n {q_i}$. Using this observation, it follows that (\ref{Ktrace}) 
vanishes for homogeneous $u$ unless $\deg (\Str_\M [(\partial D)^n u])=\deg (\partial_1W\ldots \partial_n W)-\sum_{i=1}^n {q_i}=nh-2\sum_{i=1}^n q_i$. Since 
$\psi_h(\deg \Str_\M [(\partial D)^n u])=\psi_h(\deg [(\partial D)^n u])$ when the left hand side is nonzero, we find that (\ref{Ktrace}) vanishes 
unless $\psi_h(\deg u)=\psi_h(nh-2\sum_{i=1}^n q_i)-\psi_h(\deg [(\partial D)^n])=n(\psi_h(h)-\delta_h)-\psi_h(\sum_{i=1}^n q_i)=\kappa_h$, 
where we used the relation $\psi_h(h)=2\delta_h$. 

Finally, notice that equation (\ref{Str_twist}) implies the relation:
\be
\label{LG_si}
\tr_{\M(g)}{u(g)}=(-1)^{\epsilon_h(g)}\tr_\M(u)~~,
\ee
which is equivalent with the equivariance conditions (recall the definition (\ref{gamma_def})) :
\be
\tr_{\gamma_h(g)(\M)}{\gamma_h(g)(u)}=(-1)^{\epsilon_h(g)(\epsilon_h(\kappa_h)+1)}\tr_\M(u)~~,
\ee
where we noticed that $\epsilon_h(n\delta_h)=\epsilon_h(\kappa_h)=n~(\mod ~2)$.
Setting $g=g_\tau$ or $g=g_T$, the last equation gives: 
\be
\label{Ktrace_inv}
\tr_{\tau_h(\M)}(\tau_h(u))=\tr_\M(u)~~,~~\tr_{T_h(\M)}{T_h(u)}=(-1)^{n+1}\tr_\M(T_h(u))~~.
\ee

\subsection{The graded topological field theory with shifts}

The traces (\ref{Ktrace}) descend to the total
cohomology category $\cG$, inducing maps which we denote by $\tr^\cG$.
Assuming that $\tr^\cG$ are nondegenerate, the discussion above shows
that $(\cG,G_h, \gamma^\cG_h, s^\cG_h, \tr^\cG, \kappa_h,\epsilon_h)$
is a graded open 2d TFT with shifts in the sense of Section
\ref{sec:shifts}.

Consider the triangulated category $\cT={\cal G}^0$ of graded matrix factorizations. We let $\gamma_h^\cT$ 
be the functor induced from $\gamma_h^\cG$ by restriction to morphisms of degree zero. 
As in  Section \ref{sec:gradedTFT}, $\cT$ carries traces $tr_\M^\cT:\Hom_\cT(\M,\gamma_h(\kappa)(\M))\rightarrow \C$ induced from $\cG$, 
which make it into a graded cyclic category. Equations (\ref{kappa_LG}) and (\ref{gamma_LG}) imply that the 
Serre functor $S_h=\gamma_h(\kappa_h)$ defined by this cyclic structure takes the form: 
\be
\label{S_LG}
S_h=\left\{\begin{array}{lll}
\tau_h^{\frac{h{\hat c}}{2}}\rho_h^n =T_h^n\tau_h^{\frac{h}{2}({\hat c}-n)}& {\rm ~for~} h\in 2\N~, \\
\rho_h^{h{\hat c}}=T_h^{h{\hat c}}\tau_h^{-\frac{h(h-1)}{2}{\hat c}}  &{\rm ~for~} h\in 2\N+1\ 
\end{array}\right.~~.
\ee
As in Section 2, the graded category with shifts $\cG$ can be recovered as the skew category $\cG=\cT^\bullet[G_h]$. 
Combining everything, we conclude: 

\

{\em Assume that (\ref{Ktrace}) is homologically nondegenerate, i.e. the trace $\tr^\cG$ induced on $\cG$ is nondegenerate.  
Then the system $(\cG,G_h, \gamma^\cG_h, s^\cG_h, \tr^\cG, \kappa_h,\epsilon_h)$ is a graded open
2d TFT with shifts and the corresponding system $(\cT,G_h, \gamma_h^\cT,
tr^\cT, \kappa_h,\epsilon_h)$ is an equivariant cyclic category. In
particular, (\ref{S_LG}) is an equivariant Serre functor on $\cT$ under this
assumption, when considered together with the isomorphisms $\Hom_{\cT}(\M,\N)\rightarrow \Hom_{\cT}(\N,\gamma_h(\kappa)(\M))^{\rm v}$ induced 
by $tr^\cT$. }

\

The two descriptions are related as explained in Section 2.2.
Since homological nodegeneracy of (\ref{Ktrace}) has not been proved in general, we can view (\ref{S_LG}) as a conjecture for 
the form of the Serre functor on $\cT$.  We can test this in two particular cases for 
which the Serre functor is independently known. 

\paragraph{Example 1: Minimal Landau-Ginzburg models}

Consider a minimal model at the conformal point. Then one can take $n=3$ and $B=\C[\phi_1,\phi_2,\phi_3]$. The weights $q_i$ of $\phi_i$ and 
$h$ of $W$ combine in the a regular weight vector $w:=(q_1,q_2,q_3;h)$ with: 
\be
\label{qmin}
\sum_{i=1}^{3}q_i=h+1~~. 
\ee
It is well-known that such weight vectors have an ADE classification, corresponding to the type of the 
surface singularity defined by $W$. Namely, we have
\be
\nn
w=\left\{\begin{array}{lll}
(1,b,l+1-b;l+1), b=1\ldots l  &{\rm ~for~} A_{l}\ ( l\ge 1),\\
(l-2,2,l-2; 2(l-1)) & {\rm ~for~} D_l\  ( l\ge 4),\\
(4,3, 6;12)  &{\rm ~for~} E_6,\\
(6,4,9;18)  & {\rm ~for~} E_7, \\
(10,6,15; 30)   &{\rm ~for~} E_8\ .
\end{array}\right.
\ee
and the bulk superpotential can be brought to the form:  
\be
\nn
W(x,y,z)=
\left\{\begin{array}{lll}
x^{l+1}+yz,        &{\rm ~for~} A_{l}\ ( l\ge 1),\\
x^2y+y^{l-1}+z^2, &{\rm ~for~} D_l\  ( l\ge 4),\\
x^3+y^4+z^2,     &{\rm ~for~} E_6,\\
x^3+xy^3+z^2,     &{\rm ~for~} E_7, \\
x^3+y^5+z^2,      &{\rm ~for~} E_8\ .
\end{array}\right.
\ee
In all cases, the degree $h$ of $W$ coincides with the dual Coxeter number of the ADE group. Equation (\ref{qmin}) reduces to the well-known formula: 
\be
\nn
{\hat c}=3-2\frac{h+1}{h}=1-\frac{2}{h}~~.
\ee
Using this in equation (\ref{S_LG}), we find: 
\be
\label{Smin}
S_h=T_h\tau_h^{-1}~~
\ee
for all ADE groups. In \cite{saito}, it
was shown that ${\cal T}$ admits a full and strong exceptional
collection, giving a triangle equivalence $\cT\approx
D^b(\mod_{\C Q})$, where $\C Q$ is the path algebra of the
corresponding Dynkin quiver (whose orientation can be chosen
arbitrarily). Moreover, it was shown in loc. cit that $\tau_h^{-1}$
agrees with the Auslander-Reiten translation \cite{Happel} of
$D^b(\mod_{\C Q})$.  Thus equation (\ref{Smin}) recovers the
result of \cite{reiten} relating the Serre and Auslender-Reiten
functors of a Hom-finite Krull-Schmidt category.  Also notice that
(\ref{Ttau}) becomes the known relation between the Auslender-Reiten and translation
functors of $D^b(\mod_{\C Q})$.

\paragraph{Example 2: Landau-Ginzburg models of Calabi-Yau type }

A Landau-Ginzburg model is of {\em Calabi-Yau type} if:
\be
\label{qCY}
\sum_{i=1}^{n}q_i=h~~. 
\ee
In this case, in was shown in \cite{Orlov3} that $\cT$ is triangle equivalent with $D^b(X)$, where $X$ is the 
Calabi-Yau hypersurface $W=0$ in $W\C\P^{n-1}_{q_1\ldots q_n}$. Equation (\ref{qCY}) gives: 
\be
\nn
{\hat c}=n-2=\dim X~~.
\ee
Using this in (\ref{S_LG}), we find: 
\be
S_h=T_h^{\dim X}~~,
\ee
which of course is the Serre functor of a Calabi-Yau variety.

\subsection{Component description} 

An equivalent though less natural description of $\DG_W^\gr(B)$ can be obtained by `passing to components'
as in Section \ref{sec:integrable}. For completeness, let us show explicitly how this recovers the 
formulation of \cite{Orlov3, saito}. 

As explained in Section \ref{sec:integrable}, a graded $B$-module 
of type $(G_h,\psi_h)$ can be viewed as a pair of graded $B$-modules (in our case $G_A=\Z_2$)  
which we denote by $M_+=M_{\hat 0}$ and $M_-=M_{\hat 1}$. According to the general construction, 
we have $M_+=\oplus_{\epsilon_h(g)={\hat 0}}{M^g}$, $M_-=\oplus_{\epsilon_h(g)={\hat 1}}{M^g}$, 
and the $\Z$-gradings on $M_\pm$ are specified by choosing elements $g_\pm \in G$ such that $\epsilon_h(g_+)={\hat 0}$ and 
$\epsilon_h(g_-)={\hat 1}$. Contact with \cite{Orlov3, saito} is made by choosing 
$g_+=0$ and $g_-=\delta_h$. With this choice, we have $M_\pm=\oplus_{k\in \Z}{M^k_\pm}$,
where $M_\pm^k=M^{g_\pm+\psi_h(k)}$ are given by: 
\be
\nn
M_+^k=\left\{\begin{array}{lll}
 M^{(k,{\hat 0})}& {\rm ~for~} h\in 2\N~, \\
M^{2k}&{\rm ~for~} h\in 2\N+1\ 
\end{array}\right.~~, 
~~M_-^k=\left\{\begin{array}{lll}
M^{(k+\frac{h}{2},{\hat 1})}& {\rm ~for~} h\in 2\N~, \\
M^{2k+h} &{\rm ~for~} h\in 2\N+1\ 
\end{array}\right.~~.
\ee
Thus $M=M_+\oplus M_-$ and $D$ can be decomposed as
$D=\left[\begin{array}{cc} 0 & F\\ G &0
\end{array}\right]$ with $F\in \Hom_B^{h}(M_-,M_+)$ and $G\in \Hom_B^0(M_+,M_-)$. The integrability condition $D^2=W$ becomes $FG=GF=W$.  
In this language, a homogeneous morphisms $u\in
\Hom_{\DG^\gr_W}(\M,\N)$ decomposes as $u=\left[\begin{array}{cc}
u_{++} & u_{-+}\\ u_{+-} &u_{--} \end{array}\right]$ with
$u_{\alpha\beta}\in \Hom_B(M_\alpha,N_\beta)$ and the homogeneity
conditions become:

\noindent a) When $h$ is even, we have:

a1) $u\in  \Hom_{\DG^\gr_W(B)}^{(k,{\hat 0})}(\M,\N)$ iff $u_{\pm\pm}\in \Hom^k_B(M_\pm,N_\pm)$ and $u_{\pm\mp}=0$
 
a2)  $u\in  \Hom_{\DG^\gr_W(B)}^{(k,{\hat 1})}(\M,\N)$ iff $u_{\pm\mp}\in \Hom^{k\mp \frac{h}{2}}_B(M_\pm,N_\mp)$ and $u_{\pm\pm}=0$.

\noindent b) When $h$ is odd, we have: 

b1) $u\in  \Hom_{\DG^\gr_W(B)}^{k}(\M,\N)$ with $k$ even iff $u_{\pm\pm}\in \Hom^k_B(M_\pm,N_\pm)$ and $u_{\pm\mp}=0$

b2)  $u\in  \Hom_{\DG^\gr_W(B)}^{k}(\M,\N)$ with $k$ odd iff $u_{\pm\mp}\in \Hom^{\frac{k\mp h}{2}}_B(M_\pm,N_\mp)$ and $u_{\pm\pm}=0$.

\noindent It is now clear that passing to the zeroth cohomology category $\cT$ recovers the construction of \cite{Orlov3,saito}. 

A simple case by case analysis shows that the following relations hold
for all $h$ (below $(1)$ is the usual twist functor on the category
$\gr_B$ of finitely generated $\Z$-graded modules over $B$):

(1) $\tau_h(M,D)=(M(g_\tau),D(g_\tau))$ with
$M(g_\tau)_\pm=M_\pm (1)$, $D(g_\tau)=\left[\begin{array}{cc} 0 &
F(1)\\ G(1) &0\end{array}\right]$ and $\tau_h (u)=u(g_\tau)=\left[\begin{array}{cc}
    u_{++} & u_{-+}\\ u_{+-} &u_{--}
  \end{array}\right]=\left[\begin{array}{cc} u_{++}(1) & u_{-+}(1)\\
    u_{+-}(1) &u_{--}(1) \end{array}\right]$

(2) $T_h(M,D)=(M(g_T),-D(g_T))$ with $M(g_T)_+=M_-$,~$M(g_T)_-=M_+(h)$, $D(g_T)=\left[\begin{array}{cc} 0 &
G\\ F &0\end{array}\right]$ and $T_h(u)=u(g_T)=\left[\begin{array}{cc} u_{--} & u_{+-}\\ u_{-+}
    &u_{++}(h) \end{array}\right]$.

\

\noindent 
Passing to the zeroth cohomology category $\cT=H^0(\DG_W^\gr(B))$, it follows that $\tau_h$ and $T_h$ induce the twist and shift functors considered
in \cite{Orlov3} and \cite{saito}. In this description, we have $\Str_\M(u)=\tr_{\M_+}(u_{++})-\tr_{\M_-}(u_{--})$.

\section{Conclusion} 

We discussed open 2d TFTs whose category of boundary sectors admits a
grading by an arbitrary Abelian group $G$, giving a general result
which translates between the physics-inspired formulation of such
models and the mathematical theory of Serre functors. This provides a
general method for constructing open 2d TFTs with extended grading
from more traditional mathematical data.  When the group $G$ is
sufficiently large, this equivalence explains how non-Calabi-Yau
categories arise in the general framework of \cite{CIL1, Moore}.

Using this formalism, we gave a careful discussion of grading issues
for the category of graded D-branes in B-type topological
Landau-Ginzburg models, describing its precise relation with the
triangulated category of matrix factorizations, and made a specific
proposal for an equivariant Serre functor on the latter.  Our result
shows that the category of boundary sectors can be recovered as the
graded skew category of the triangulated category of \cite{Orlov3}
with respect to the Abelian group generated by the shift and twist
functors of \cite{Orlov3, saito}. This group depends on the parity of
the bulk superpotential and recovers the correct grading on the spaces
of boundary and boundary condition changing observables. We also gave
a description of the dG category of graded D-branes which manifestly
displays the grading by characters of the full vector-axial R-symmetry preserved by 
such models.

\appendix

\section{Equivariant Serre functors.}
\label{sec:Serre}

Let us fix a (not necessarily Abelian) group $G$.
Consider a $G$-category $(\cT,G,\gamma)$ and set $G^*\:=\Hom(G,\C^*)$. 
It is well-known that Serre functors on $\cT$ can be described by pairs $(S,tr)$ where 
$S$ is an automorphism\footnote{One can extend this to autoequivalences, but we restrict to automorphisms for simplicity.} 
of $\cT$ and $\tr$ is a family of linear maps $tr_a:\Hom_\cT(a,S(a))\rightarrow \C$ define for all $a\in \Ob\cT$, 
subject to the conditions: 
\be
\nn
tr_a(vu)=tr_b(S(u)v) ~~\forall u\in \Hom_\cT(a,b)~~\forall v\in \Hom_\cT(b,S(a))~~
\ee
and: 
\be
\nn
tr_a(vu)=0~~\forall u\in \Hom_\cT(b,S(a))\Rightarrow u=0~~.
\ee

A Serre functor
$(S,tr)$ on $\cT$ is called {\em $G$-equivariant} if $S$ is
$G$-equivariant and there exists a framing
$\eta(g):S\circ \gamma(g)\stackrel{\sim}{\rightarrow} \gamma(g) \circ S$ of $S$ and a character $\chi \in G^*$ which
satisfy the conditions:
\be
\label{S_eq}
tr_{\gamma(g)(a)}{\gamma(g)(u)}=\chi(g) tr_a(\gamma(g)^{-1}(\eta_a(g))\circ u)~~\forall a\in \Ob\cT~~\forall u\in 
\Hom_{\cT}(a, (\gamma(g)^{-1}S\gamma(g))(a))
\ee
for all $g\in G$. 

In the situation of Subsection 2.2, we have $G=$ Abelian, $S=$ $G$-invariant, $\eta=\id=$ trivial framing 
and $\chi(g)=(-1)^{\epsilon(g)(\epsilon(\kappa)+1)}$ for some $\kappa\in G$.

\section{Worldsheet analysis of R-symmetries in LG models}

\label{sec:worldsheet}

Consider  the B-twisted topological Landau-Ginzburg model with target $\C^n$, whose boundary
coupling was given in \cite{KY, Lerche_LG} for the simplest D-branes
and constructed for arbitrary branes in \cite{LG1, LG2}. In
the set-up of the latter papers, the untwisted model has
worldsheet bosons $\phi^i$ and $\phi^{\bar i}$ corresponding to local
complex coordinates of a non-compact target Calabi-Yau manifold. It also has
worldsheet fermions $\psi^i_+,\psi^{\bar i}_+$ which are
sections of $K^\frac{1}{2}\otimes \phi^*(T^{1,0}X)$ and
$K^\frac{1}{2}\otimes \phi^*(T^{0,1}X)$, as well as 
$\psi^i_-,\psi^{\bar i}_-$ which are sections of ${\bar
K}^\frac{1}{2}\otimes \phi^*(T^{1,0}X)$ and ${\bar
K}^\frac{1}{2}\otimes \phi^*(T^{0,1}X)$. Here $K$ and ${\bar K}$ are
the canonical line bundle on the worldsheet and the bundle of $(0,1)$
forms. Performing the B-twist as in \cite{Witten_mirror} replaces
$\psi^i_+$ with a section $\rho^i_+$ of $K\otimes \phi^*(T^{1,0}X)$
and $\psi^i_-$ with a section $\rho^i_-$ of ${\bar K}\otimes
\phi^*(T^{1,0}X)$. It also replaces $\psi^{\bar i}_+$ with a section
$\chi^{\bar i}$ of $\phi^*(T^{0,1}X)$ and $\psi^{\bar i}_-$ with a
section ${\bar \chi}^{\bar i}$ of the same bundle
$\phi^*(T^{0,1}X)$. The later combine into the new Grassmann odd
fields $\eta^{\bar i}=\chi^{\bar i}+{\bar \chi}^{\bar i}$ and
$\theta_i=G_{i{\bar j}}(\chi^{\bar j}-{\bar \chi^{\bar j}})$.  Taking
the bulk superpotential $W$ to be homogeneous\footnote{Due to the
superspace formulation, it is traditional to take $W$ to formally have
degree $2$ by using the `fractional charges' ${\tilde q}_i:=2q_i/h$
instead of $q_i$. Here we prefer to work with integral charges since
they are directly related to the characters of $U(1)$.} of degree $h$:
\be
\nn
W(\{e^{2is q_i}\phi_i\})=e^{2ish} W(\{\phi_i\})~~,
\ee
the model has a vector $U(1)$ R-symmetry which acts as:
\bea
&& \phi^i\rightarrow e^{2isq_i}\phi^i~~,~~\phi^{\bar i}\rightarrow e^{-2isq_i}\phi^{\bar i}~~,\nn\\
&& \psi_\pm^i\rightarrow e^{is (2q_i-h)}\psi_\pm^i~~,~~\psi_\pm^{\bar i}\rightarrow e^{-is (2q_i-h)}\psi_\pm^{\bar i}
\eea
as well as an axial $U(1)$ R-symmetry given by: 
\bea
&& \phi^i\rightarrow \phi^i~~,~~\phi^{\bar i}\rightarrow \phi^{\bar i}~~,\nn\\
&& \psi_\pm^i\rightarrow e^{\mp 2it}\psi_\pm^i~~,~~\psi_\pm^{\bar i}\rightarrow e^{\pm 2it }\psi_\pm^{\bar i}~~,
\eea
The $U(1)_A$ group is canonically parameterized by $e^{2it}$ with $t\in [0,\pi]$, while the range of $s$ depends on the 
parity of $h$. For even $h$, we take $s\in [0,\pi]$, while for odd $h$ we take $s\in [0,2\pi]$. This insures that 
the action of the vector R-symmetry group is a proper representation. Notice that for odd $h$, the $U(1)_V$ group is 
the double cover of the circle $\{e^{2is}|s\in [0,\pi]\}$; of course, the double cover is again a circle group.
These actions translate as follows for the Grassmann-odd fields of the twisted theory: 
\be
\label{UV}
\rho^i\rightarrow e^{is (2q_i-h)}\rho^i~~,~~\eta^{\bar i}\rightarrow e^{-is (2q_i-h)}~\eta^{\bar i}~~,~~
\theta_i\rightarrow e^{-is (2q_i-h)}~\theta^{\bar i}
\ee
for $U(1)_V$ and: 
\be
\label{UA}
\rho^i_\pm\rightarrow e^{\mp 2it}\rho^i_\pm,~~,~~\chi^{\bar i}\rightarrow e^{+2it}~\chi^{\bar i}~~,~~
{\bar \chi}_i\rightarrow e^{-2it} {\bar \chi}^{\bar i}~~
\ee
for $U(1)_A$. As explained in \cite{LG1,LG2}, a general topological
D-brane of this model is described by a superbundle $E=E_+\oplus E_-$
over the target space, together with a superconnection $\cA$ whose
super-curvature can be taken to satisfy ${\cal F}^{(0,\leq 2)}=W$. The
boundary coupling is given by the super-Wilson loops:
\be
\label{calU}
{\cal U}:=\Str Pe^{-\oint_{C}{d\tau {\cal M}}}~~.
\ee
where $C$ is a circle boundary of the worldsheet. The quantity ${\cal M}$ is given by:
\be
\label{M}
{\cal M}=\left[\ba{cc} 
{\hat {\cal A}}^{(+)}+\frac{i}{2}(FF^\dagger+G^\dagger G) &
\frac{1}{2}\rho^i_0 \nabla_i F+\frac{i}{2}\eta^{\bar i}\nabla_{\bar i}G^\dagger
\\
\frac{1}{2} \rho^i_0 \nabla_i G +\frac{i}{2}\eta^{\bar i}\nabla_{\bar
  i}F^\dagger &
{\hat {\cal A}}^{(-)} +\frac{i}{2}(F^\dagger F+GG^\dagger)
\ea\right]~~,
\ee
where $\rho_0^i d\tau_a$ is the pull-back of $\rho^i$ to $C_a$ and:
\be
\nn
{\hat {\cal A}}^{(\pm)}:=A_{\bar i}^{(\pm)}{\dot \phi}^{\bar i} +\frac{1}{2}\eta^{\bar i}  
F^{(\pm)}_{{\bar i} j}\rho_0^j
\ee
are connections on the bundles ${\cal E}_\pm$ obtained by pulling back
$E_\pm$ to the boundary. The dot stands for the
derivative $\frac{d}{d\tau}$. One has ${\cal M}={\hat {\cal A}}+\Delta+K$,
where: 
\be
\nn
\Delta:=\frac{1}{2}\rho_0^i\partial_i D~~,
\ee
\be
\nn
K:=\frac{i}{2}\left(\eta^{\bar i}\nabla_{\bar i}D^\dagger +[D,D^\dagger]_+ \right)
\ee
and:
\be
\nn
{\hat {\cal A}}={\dot \phi}^{\bar i}A_{\bar i}+\frac{1}{2}F_{{\bar
    i}j}\eta^{\bar i}\rho^j_0~~.
\ee
Here $A=A_id\phi^i+A_{\bar i}d\phi^{\bar i}=A^{(+)}+A^{(-)}$ is the direct sum
connection on $\End(E)$ induced by the 1-form part of $\cA$, while
$D=\left[\begin{array}{cc} 0 & F\\ G &0 \end{array}\right]\in
\End^{\rm odd}(E)$ is the zero-form part of $\cA$, which satisfies
$D^2=W$ and plays the role of topological tachyon condensate. 

Consider the coordinate ring $B=\C[\phi_1\ldots \phi_n]$ of the target
space $\C^n$.  Passing to holomorphic sections of $E$, one finds
\cite{LG1,LG2} that boundary conditions correspond to pairs $\M=(M,D)$
where $M$ is a free finitely-generated
$B$-supermodule and $D\in
\End_B^{\rm odd}(M)$ satisfies the constraint $D^2=W$. We have a
$\Z_2$-graded dG category $\DG_W(B)$ whose objects are pairs of this form
and whose morphisms are $\Hom_{\DG_W(B)}(\M,\N):=\Hom_B(M,N)$, endowed
with the differential $d_{\M,\N}(u)=D_N\circ u -(-1)^{|u|}u\circ D_M$,
where $|u|\in \Z_2$ denotes the $\Z_2$-degree of $u$. This is the dG category
of all topological D-branes.  The space of boundary
observables reduces to $H_{d_{\M\N}}(\Hom_B(M,N))$.

The axial R-symmetry is broken by the bulk superpotential to the
subgroup $\Gamma_A\approx \Z_2$ whose generator corresponds to $t=\frac{\pi}{2}$
in (\ref{UA}), giving the residual $\Z_2$ action:
\be
\nn
\rho^i\rightarrow -\rho^i~~,~~\eta^{\bar i}\rightarrow -\eta^{\bar i}~~,~~
\theta_i\rightarrow -\theta^{\bar i}~~.
\ee
One finds that the boundary coupling
is invariant if one takes the tachyon condensate to transform as:
\be
\nn
D\rightarrow U_A D U_A^{-1}=-D~~,
\ee
where $U_A=\left[\begin{array}{cc} 1 & 0\\ 0 & -1
\end{array}\right]$ satisfies $U_A^2=1$.  This shows
that we must take the residual axial R-symmetry to act on $\Hom_B(M,N)$ as
$u\rightarrow U_A u U_A^{-1}$, i.e.: \be \nn u\rightarrow (-1)^{|u|}u~~\forall ~~{\rm homogeneous}~~~~u\in \Hom_B(M,N)~~.  \ee Hence the
$\Z_2$-grading of $\DG_W(B)$ is induced by
this residual symmetry. 

The connections $A^{(\pm)}$ are $U(1)_V$-invariant, which amounts to the
component transformation laws $A_i^{(\pm)}(\{e^{2is q_i}\phi_i\})=e^{-2is q_i}
A_i^{(\pm)}(\{\phi_i\})$ and $A_{\bar i}^{(\pm)}(\{e^{2is q_i}\phi_i\})=e^{+2is q_i}
A_{\bar i}^{(\pm)}(\{\phi_i\})$. It is easy to check that the boundary coupling
is $U(1)_V$-invariant provided that $F$ and $G$ satisfy:
\be
\nn
F(\{e^{2is q_i}\phi_i\})=e^{2ish}F(\{\phi_i\})~~,~~G(\{e^{2is q_i}\phi_i\})=G(\{\phi_i\})~~.
\ee
Indeed, this corresponds to $U_V(s)D(\{e^{2isq_i}\phi_i\})U_V(s)^{-1}=e^{ish}D(\{\phi_i\})$, 
where 
$U_V(s):=\left[\begin{array}{cc} 1 & 0\\ 0 & e^{ish}
\end{array}\right]$ gives a representation of $U(1)_V$.
Then (\ref{M}) satisfies ${\cal M}(\{e^{2is
q_i}\phi_i\})=U_V(s)^{-1}{\cal M}(\{\phi_i\}) 
U_V(s)$ so (\ref{calU}) is invariant. We have $\Gamma_V\approx U(1)$ and $\Gamma_V^*\approx U(1)^*\approx
\Z$. 

To take the vector R-symmetry into account, we grade the ring $B$
by associating weights $q_i$ to $\phi_i$.  The analysis of \cite{Walcher} shows
that $U(1)_V$-invariant boundary conditions correspond to those pairs
$\M=(M,D)\in \Ob \DG_W$ for which $M_\pm$ are free graded $B$-modules
and $D=\left[\begin{array}{cc} 0 & F\\ G &0 \end{array}\right]$ is such
that $F$ and $G$ are homogeneous of degrees $h$ and $0$ respectively.
We will refer to this grading as the `vector grading'. The dG category
of `graded branes' is the full subcategory $\DG_W^\gr(B)$ of $\DG_W(B)$
consisting of such pairs; this is the dG category of those topological
D-branes whose boundary conditions preserve the full $U(1)_V$
symmetry.  For any objects $\M$ and $\N$ of $\DG_W^\gr$, we have 
an action of $U(1)_V$ on $\Hom_B(\M,\N)$ given by:
\be
\label{Vaction}
u\rightarrow U_V(s) u(\{e^{2isq_i}\phi_i\})U_V(s)^{-1}~~,
\ee
as well as the action of the axial $\Z_2$ symmetry: 
\be
\label{Aaction}
u\rightarrow U_A u U_A^{-1}~~.
\ee
Since $[U_A,U_V(s)]=0$, these combine into an action of the 
group $\Gamma=\Gamma_V\times \Gamma_A\approx U(1)\times \Z_2$.

The morphism $u\in \Hom_{\DG_W^\gr(B)}(\M,\N)$ is `vector homogeneous' of degree $q\in \frac{1}{2}\Z$ 
if it satisfies:
\be
\nn
U_V(s) u(\{e^{2isq_i}\phi_i\})U_V(s)^{-1}=e^{2isq} u(\{\phi_i\})~~.
\ee
With this definition, the condensate $D$ of every object $\M=(M,D)$ of $\DG_W^\gr(B)$ is homogeneous 
of vector degree $\frac{h}{2}$. 
Writing $u=\left[\begin{array}{cc} u_{++} & u_{-+}\\ u_{+-} &u_{--} \end{array}\right]$ 
with $u_{\alpha\beta}\in \Hom_B(M_\alpha,N_\beta)$, this condition amounts to:
\be
\nn
\deg u_{++}=\deg u_{--}=q~~,~~\deg u_{-+}=q+\frac{h}{2}~~,~~\deg u_{+-}=q-\frac{h}{2}~~.
\ee
Since $u_{\alpha\beta}$ must have integral degrees, this requires $q\in \Z$ for $u\in \Hom_{\DG_W^\gr(B)}^{{\hat 0}}(\M,\N)$
and $q\in \Z+\frac{h}{2}$ for $u\in \Hom_{\DG_W^\gr(B)}^{{\hat 1}}(\M,\N)$. We find the decompositions: 
\bea
\Hom_{\DG_W^\gr(B)}^{{\hat 0}}(\M,\N)&=&\oplus_{q\in \Z}{\Hom_{\DG_W^\gr(B)}^{[q, {\hat 0}]}(\M,\N)}\nn\\
\Hom_{\DG_W^\gr (B)}^{{\hat 1}}(\M,\N)&=&\oplus_{q\in \Z+\frac{h}{2}}{\Hom_{\DG_W^\gr(B)}^{[q,{\hat 1}]}(\M,\N)}~~.\nn
\eea
We have $D_\M\in \Hom_{\DG_W^\gr(B)}^{[\frac{h}{2}, {\hat 1}]}(\M,\M)$, so the differential $d_{\M,\N}$ satisfies: 
\be
\label{dhom}
d_{\M,\N}(\Hom_{\DG_W^\gr (B)}^{[q, \alpha]}(\M,\N))\subset \Hom_{\DG_W^\gr (B)}^{[q+\frac{h}{2}, \alpha+{\hat 1}]}(\M,\N)~~.
\ee
When $h$ is even, we have $\Z+\frac{h}{2}=\Z$ and the decompositions above give a $\Z\times \Z_2$-grading: 
\be
\nn
\Hom_{\DG_W^\gr (B)}(\M,\N)=\oplus_{ q\in \Z, \alpha\in \Z_2}{\Hom_{\DG_W^\gr (B)}^{[q, {\hat \alpha}]}(\M,\N)}~~.
\ee
When $h$ is odd, we have $\Z+\frac{h}{2}=(\frac{1}{2}\Z)\setminus \Z$ and we find a $\frac{1}{2}\Z$-grading: 
\be
\nn
\Hom_{\DG_W^\gr (B)}(\M,\N)=\oplus_{q\in \frac{1}{2}\Z}{\Hom_{\DG_W^\gr (B)}^{[q]}(\M,\N)}~~,
\ee
where $\Hom_{\DG_W^\gr (B)}^{[q]}(\M,\N):=\Hom_{\DG_W^\gr (B)}^{[q,{\hat
0}]}(\M,\N)$ if $q\in \Z$ and
$\Hom_{\DG_W^\gr (B)}^{[q]}(\M,\N):=\Hom_{\DG_W^\gr (B)}^{[q,{\hat 1}]}(\M,\N)$ if
$q\in (\frac{1}{2}\Z)\setminus \Z$.  

It is easy to see that both of these decompositions are compatible
with the composition of morphisms.  Equation (\ref{dhom}) shows that $d_{\M\N}$
are homogeneous of bidegree $(\frac{h}{2},{\hat 1})\in \Z\times \Z_2$
when $h$ is even, and of degree $\frac{h}{2}\in \Z$ when $h$ is
odd. Setting:
\be
\nn
G_h=\left\{\begin{array}{lll}
\Z\times \Z_2& {\rm ~for~} h\in 2\N~, \\
\Z  &{\rm ~for~} h\in 2\N+1\ 
\end{array}\right.~~,
\ee
as in (\ref{G_h}), we redefine the grading for odd $h$ so that all degrees 
become integral. Thus we set $\Hom_{\DG_W^\gr (B)}(\M,\N)=\oplus_{g\in G_h} \Hom_{\DG_W^\gr (B)}^{g}(\M,\N)$, where 
\be
\label{Hom_grading}
\Hom_{\DG_W^\gr (B)}^{g}(\M,\N)=
\left\{\begin{array}{lll}
\Hom_{\DG_W^\gr (B)}^{[k,\alpha]}(\M,\N) & {\rm ~for~} h\in 2\N {\rm ~and~} g=(k,\alpha)\in \Z\times \Z_2~~, \\
\Hom_{\DG_W^\gr (B)}^{[k/2]}(\M,\N)  &{\rm ~for~} h\in 2\N+1 {\rm ~and~} g=k\in \Z \ 
\end{array}\right.~~.
\ee
With this convention, $\DG_W^\gr (B)$ is a $G_h$-graded category. We let $\deg u\in G_h$ denote the $G_h$-degree of morphisms, 
which is given by:
\be
\nn
\deg u =\left\{\begin{array}{lll}
(q,\alpha)\in \Z\times \Z_2& {\rm ~for~} h\in 2\N, \\
2q \in \Z  &{\rm ~for~} h\in 2\N+1\ 
\end{array}\right.
\ee
Notice that the rescaled vector R-charge of $u$ is recovered as $\phi_h(\deg u)$, where $\phi_h:G\rightarrow \Z$ is the group 
morphism given by: 
\be
\nn
\left\{\begin{array}{lll}
\phi_h(k,\alpha)=k & {\rm ~for~} h\in 2\N, \\
\phi_h(k)=k &{\rm ~for~} h\in 2\N+1\ 
\end{array}\right.~~.
\ee
The true vector R-charge (which is valued in $\frac{1}{2}\Z$) is given by: 
\be
\nn
q(u)=\left\{\begin{array}{lll}
\phi_h(\deg u) & {\rm ~for~} h\in 2\N, \\
\frac{1}{2}\phi_h(\deg u) &{\rm ~for~} h\in 2\N+1\ 
\end{array}\right.~~.
\ee 
On the other hand, the axial charge is recovered as $\epsilon_h(\deg u)$, where $\epsilon_h:G_h\rightarrow \Z_2$
is the group morphism given in (\ref{epsilon_h}).

The above can also be described as follows. We have a decomposition:
\be
\label{M_dec}
M=\oplus_{q\in \Z}M^{[q,{\hat 0}]}\oplus \oplus_{q\in \Z+\frac{h}{2}}M^{[q,{\hat 1}]}~~,
\ee
with $M^{[q,{\hat 0}]}=M_+^q$ and $M^{[q,{\hat 1}]}=M^{q-\frac{h}{2}}_-$, where $M_\pm=\oplus_{k\in \Z}M_\pm^k$ are the decompositions 
of $M_\pm$ as graded $B$-modules.  It is clear that the components satisfy $M^{[q,\alpha]}B^k\subset M^{[q+k,\alpha]}$.

For even $h$, equation (\ref{M_dec}) gives a $\Z\times \Z_2$ grading
$M=\oplus_{q\in \Z, \alpha\in \Z_2}M^{[q,\alpha]}$, while for odd $h$
it gives a $\frac{1}{2}\Z$-grading $M=\oplus_{q\in \frac{1}{2}\Z
}M^{[q]}$, where $M^{[q]}:=M^{[q,{\hat 0}]}$ for $q\in \Z$ and
$M^{[q]}:=M^{[q,{\hat 1}]}$ for $q\in \Z+\frac{1}{2}$. Rescaling $q$
for odd $h$ such that all charges become integral, we find that the
underlying vector space of $M$ is $G_h$-graded with components
$M^{(q,\alpha)}=M^{[q,\alpha]}$ for $h$ even, and $M^{q}:=M^{[q/2]}$
for $h$ odd. This rescaled grading satisfies $M^gB^k\subset
M^{g+\psi_h(k)}$, where the morphism $\psi_h:\Z\rightarrow G$ is given
in (\ref{psi_h}). Hence the objects of $\DG_W^\gr(B)$ are integrable
modules over the curved differential graded algebra $(B,0,W)$, as
discussed in Section 4. The grading (\ref{Hom_grading}) on $\DG_W^\gr
(B)$ is naturally induced by these $G_h$-gradings on objects.

Finally, we notice that the difference made by the parity of $h$ can
also be seen directly.  The group $\Gamma=\Gamma_V\times
\Gamma_A=U(1)\times \Z_2$ acts on $\Hom_{\DG^\gr_W (B)}(\M,\N)$ by the
product of the actions (\ref{Vaction}) and (\ref{Aaction}).  This
combined action is faithful for generic $\M,\N$ when $h$ is even, but
has a $\Z_2$ kernel when $h$ is odd (in the the latter case we have
$U_V(\pi)=U_A$). Hence the `effective' symmetry group equals
$U(1)\times \Z_2$ when $h$ is even and $[U(1)\times \Z_2]/\Z_2\approx
U(1)$ when $h$ is odd.  As a consequence, its group of characters
equals $\Z\times \Z_2$ or $\Z$ respectively. Of course, this is the
group $G_h$ considered above.

\end{document}